\documentclass[journal,onecolumn,11pt]{IEEEtran}
\usepackage{amsmath,theorem,amssymb,graphicx,amsfonts,epsfig,cite,subfigure,times,latexsym,array,verbatim,url}

\newcommand{\emb}[1]{\text{\boldmath{$#1$}}}

\newcommand{\indic}[1]{\mbox{$1\!\!1$}{\{#1\}}}

\theorembodyfont{\itshape}
\newtheorem{theorem}{Theorem}[section]
\newtheorem{lemma}{Lemma}[section]
\newtheorem{proposition}{Proposition}[section]

\newtheorem{remark}{Remark}[section]
\graphicspath{{figures/}}
\def \r{\rho}
\def \br{b_{\rho}}
\def \fr{f_{\rho}}
\def \gr{g_{\rho}}
\def \ur{u_{\rho}}
\def \cs{{\cal S}}
\def \om{\omega}
\def \Om1i{\Omega^1_i}
\def \om0c{\omega^c_0}

\begin{document}
\title{Boolean Compressed Sensing and\\ Noisy Group Testing\thanks{This research was partially supported by NSF CAREER award ECS 0449194
and NSF Grant 0932114, by the U.S. Department of Homeland Security
under Award Number 2008-ST-061-ED0001}}
\author{George Kamal Atia$^\dagger$,~\IEEEmembership{Member,~IEEE}, and Venkatesh Saligrama$^\ddagger$,~\IEEEmembership{Senior Member,~IEEE}
\thanks{$^\dagger$ G. Atia is with the Coordinated Science Laboratory (CSL) at the Department of Electrical and Computer Engineering at the University of Illinois at Urbana-Champaign.

$^\ddagger$ V. Saligrama is with the Information Systems and Sciences Laboratory (ISS) at the Department of Electrical and Computer Engineering at Boston University.

This work was presented in part at the Allerton Conference on Communication, Control, and Computing, Monticello IL, in September 2009 \cite{allerton_version}. The authors can be reached at: atia1@illinois.edu, srv@bu.edu.}}
\maketitle

\begin{abstract}
The fundamental task of group testing is to recover a small distinguished subset of items from a large population while efficiently reducing the total number of tests (measurements). The key contribution of this paper is in adopting a new information-theoretic perspective on group testing problems. We formulate the group testing problem as a channel coding/decoding problem and derive a single-letter characterization for the total number of tests used to identify the defective set. Although the focus of this paper is primarily on group testing, our main result is generally applicable to other compressive sensing models.

The single letter characterization is shown to be order-wise tight for many interesting noisy group testing scenarios. Specifically, we consider an additive Bernoulli($q$) noise model where we show that, for $N$ items and $K$ defectives, the number of tests $T$ is $O\left(\frac{K\log N}{1-q}\right)$ for arbitrarily small average error probability and $O\left(\frac{K^2\log N}{1-q}\right)$ for a worst case error criterion. We also consider dilution effects whereby a defective item in a positive pool might get diluted with probability $u$ and potentially missed. In this case, it is shown that $T$ is $O\left(\frac{K\log N}{(1-u)^2}\right)$ and $O\left(\frac{K^2\log N}{(1-u)^2}\right)$ for the average and the worst case error criteria, respectively. Furthermore, our bounds allow us to verify existing known bounds for noiseless group testing including the deterministic noise-free case and approximate reconstruction with bounded distortion. Our proof of achievability is based on random coding and the analysis of a Maximum Likelihood Detector, and our information theoretic lower bound is based on Fano's inequality.
\end{abstract}

\section{Introduction}
Group testing has been effectively used in numerous applications. It was originally proposed during World War II to reduce the total number of blood tests required to detect soldiers with Syphilis \cite{du,dorfman}. Instead of conducting a separate blood test for each and every soldier, the idea was to pool blood samples from many soldiers and test them simultaneously.
Group testing has been used in general in biology for screening libraries of DNA clones (strings of DNA sequence) of the human genome and for screening blood for diseases. Other applications include quality control for detecting defective parts in production lines, data forensics to test collections of documents by applying one-way hash functions, computer fault diagnosis, and contention algorithms in multiple access communications. Recently, group testing methods have also been applied to spectrum enforcement in cognitive radios ~\cite{dyspan08,Asilomar08}.

In the basic group testing problem~\cite{du}, we are given a population of $N$ items. Among them, at most $K$ items, also called defectives, are of interest. The set of defectives is denoted by $\cal G$.
Associated with the group testing problem is a binary matrix $\emb{X}$ known as the measurement matrix. This matrix defines the assignment of each of the items to different pools or collections. The $(i,j)$ entry is $1$ if the $i$-th item is contained in the $j$-th pool and $0$ otherwise. A test conducted on the pool is positive if there is at least one item belonging to the pool which is also an element of $\cal G$, and is negative otherwise.
A measurement matrix for a non-adaptive group testing algorithm is a $N \times T$ matrix where the $T$ columns correspond to pools of items and $N$ is the number of items. Each item is associated with a row codeword of length $T$. If $K$ items are defective, then the $T$ tests are a boolean sum of $K$ rows of the measurement matrix. The goal is to construct a pooling design to recover the defective set while reducing the required number of tests. Group testing is related in spirit to compressed sensing (CS). In CS we are given an N-dimensional sparse signal with support size $K$. Random projections of the sparse signal are obtained. The goal is to identify the support set while minimizing the number of projections. In this sense, group testing can be viewed as a boolean version of CS where we apply a measurement matrix to a sparse vector corresponding to the defective set with the goal of reconstructing the support, i.e. identify the defective items. The main distinction is that in group testing the test matrix used to collect the measurements is binary, so is the sparse vector, and the arithmetic is boolean. Multiplication of $0$s and $1$s is the logical AND (coincides with usual multiplication) but addition is replaced by the logical OR. 

While the degradation of CS with noise has been characterized (see ~\cite{donoho,candes,wainwright1,wainwright2,gastpar,goyal,fletcher} and references therein), the noisy group testing problem, the main focus of this paper, has been mostly unexplored. Our attention in this paper is on the so called non-adaptive testing problem~\cite{du}, where the measurement matrix is formed prior to performing the tests.

A significant part of the existing research on group testing is focused on combinatorial pool design (i.e. construction of measurement matrices) to guarantee the detection of the items of interest using a small number of tests. Two types of matrix constructions have been considered. Disjunct matrices~\cite{du} satisfy the so called covering property\footnote{We say that a row $x$ is covered by a row $y$ iff $x\vee y=y$.}. In the context of group testing this property implies that a test pattern obtained by taking any $K$ rows of the measurement matrix does not cover any other boolean sum of $K$ or smaller number of rows\footnote{In the combinatorial pool design literature, the roles of rows and columns are interchanged. In other words, the measurement matrix is $\emb{X}^T$, i.e., the transpose of our matrix $\emb{X}$. What motivates our choice is consistency with the standard information theoretic convention for codewords being rows of a codebook.}. Equivalently, for any $K+1$ rows, there always exists a column with $1$ in a row and $0$s in the other $K$ rows. Matrices that satisfy this property are often referred to as superimposed codes and combinatorial constructions were extensively developed by ~\cite{kautz,dyachkov,erdos}. Superimposed codes are not only desirable because they ensure identifiability but they also lead to efficient decoding. Separability~\cite{du} is a weaker notion that is also often employed. A separable matrix ensures that the boolean sums of $K$ rows are all distinct, which ensures identifiability. Uniquely decipherable codes~\cite{dyachkov,kautz} are codes that guarantee that every boolean sum of $K$ or smaller number of rows are distinct. Recently, it has been shown~\cite{chen} that all of these notions are equivalent up to a scaling factor on the number of tests $T$.

A different approach to group testing based on probabilistic method has also been advocated by several researchers~\cite{rashad,ruszinko,sebo,berger,gilbert}. Dyachkov and Rykov \cite{rashad}, Du and Hwang \cite{duhwang}, and Ruszinko \cite{ruszinko} developed upper and lower bounds on the number of rows $T$ for a matrix to be $K$-disjunct (bound on length of superimposed codes). Random designs were used to compute upper bounds on the lengths of superimposed codes by investigating when randomly generated matrices have the desired covering/separability properties. They showed that for $N\rightarrow\infty$ and $K\rightarrow\infty$, the number of tests $T$ must scale as $T=O\left(\frac{K^2\log N}{\log K}\right)$ for exact reconstruction with worst-case input.
Sebo \cite{sebo} investigated average error probabilities and showed that for an arbitrarily small error probability, a randomly generated matrix will be $K$-disjunct if $T=O(K\log N)$ as $N\rightarrow\infty$. Recently Berger et al. \cite{berger} proved upper and lower bounds on the number of tests for two-stage disjunctive testing. Approximate reconstruction \cite{gilbert}, whereby a fraction $\alpha$ of the defective items are allowed to be in error, has been described. Again the number of tests here has been shown to scale as $T=O(K\log N)$ as $N\rightarrow\infty$.

While these approaches have generally characterized fundamental tradeoffs for noiseless group testing, the noisy counterpart of group testing has been largely unexplored.
In this paper we present a novel information theoretic approach to group testing problems. The common approach to previous related work was to prove bounds on the size of randomly generated matrices to exhibit the aforementioned separability and covering properties. In contrast, we formulate the problem as a detection problem and establish its connection to Shannon coding theory ~\cite{coverbook}. While there exists a one-to-one mapping between both formulations, the new perspective allows us to easily obtain results for a wide range of models including noisy versions of group testing. Our approach, which is fairly general, is to map the group testing problem to a corresponding channel model which allows the computation of simple mutual information expressions to derive achievable bounds on the required number of tests. Very recently, it came to our attention that related work that uses a similar information theoretic approach for noisy group testing has been independently explored in the Russian literature \cite{malyutov,malyutov1,malyutov2,malyutov3,dyachkov_lectures}. One main difference between those approaches and our approach is that, in the earlier work, the number of defective items, $K$, is held fixed while the number of items, $N$, approaches infinity. Consequently, those bounds suggest that the number of tests must scale poly-logarithmically in $N$ regardless of $K$ for error probability to approach zero. In contrast we consider the fully high-dimensional setting wherein both the number of defectives as well as the number of items can approach infinity. This implies that our bounds not only depend on $N$ but also explicitly on the number of defectives. In addition our approach also has other advantages, including:
\begin{itemize}
\item {\bf Mutual information characterization:} Our main result is a simple single letter characterization providing order-wise tight necessary and sufficient conditions on the total number of tests. 
\item {\bf Characterization for new group testing problems:} This result allows us to verify existing bounds for some of the known scenarios and extend the analysis to many new interesting setups including noisy versions of group testing. We present bounds that explicitly characterizes the required number of tests in terms of the number of defectives and the number of items. Our analysis includes the fully high-dimensional case where both the number of defectives and the number of items approach infinity.
\item {\bf Extensions to sparse models:} Although the focus of this paper is on the Boolean case, the main result and the methods that we develop in this paper are more generally applicable to Compressed Sensing models and not just to group testing. The results extend beyond binary variables and Boolean arithmetic to the general problem of support recovery with arbitrary discrete alphabet \cite{icalp}. Consequently, some of the information theoretic Compressed Sensing results \cite{shuchin} can also be recovered using the mutual information expressions derived herein.
\end{itemize}
One major contribution of this paper is to develop tools to analyze long standing noisy versions of the group testing problem. In particular we consider two models: The dilution model and the additive model.

\begin{itemize}
\item {\bf Additive model:} False alarms could arise from errors in some of the screening tests. This arises when some tests are erroneously positive.

\item {\bf Dilution model:} Even though a positive item is contained in a given pool, the test's outcome could be negative if the defective item gets diluted for that specific test. For example, in blood testing the positive sample might get diluted in one or more tests leading to potential misses of infected blood samples. To account for such a case we analyze the group testing problem where some of the positive entries get flipped into zeros with a given probability $u$.
\end{itemize}

We also consider the case of partial reconstruction where we only aim for approximate recovery of the defective set. An error occurs only if the number of missed defectives is greater than a fraction $\alpha$ of the total $K$ defectives.

\begin{table}
\centering
    \begin{tabular}{|c|c|c|}
    \hline
    Model & $T$ (average $P_e$ criterion) & $T$ ($\max P_e$ criterion)\\ \hline
    Noise-free& $O(K\log N)$ & $O(K^2\log N)$ \\ \hline
    Partial reconstruction & $O(K\log N)$ & $O(K\log N)$ \\ \hline
    Noisy with additive noise & $O\left(\frac{K\log N}{1-q}\right)$ & $O\left(\frac{K^2\log N}{1-q}\right)$\\ \hline
    Noisy with dilution & $O\left(\frac{K \log N}{(1-u)^2}\right)$ & $O\left(\frac{K^2 \log N}{(1-u)^2}\right)$\\
    \hline
   \end{tabular}
  \caption{This table summarizes the scaling results for the various models considered in the paper which hold asymptotically for $N\rightarrow\infty$. In particular, it shows the required number of tests $T$ as a function of the size of the defective set $K$, the total number of items $N$, and the model parameters for the noiseless, additive noise, and dilution models for both the average error and worst-case error criteria. The probability $q$ is the parameter of the Bernoulli($q$) additive noise, and $u$ is the dilution probability for the dilution model.}
  \label{tab:summary}
\end{table}
Table \ref{tab:summary} summarizes the scaling results we obtained
for the various models considered in the paper for the average error
and worst case error criteria.

The rest of the paper is organized as follows. Section
\ref{sec,problem setup} describes the problem setup. The main
achievable result mapping the problem to mutual information
expression is provided in Section \ref{sec,main_result}. In
Section \ref{sec,LB} we prove a converse bound using Fano's
inequality \cite{coverbook}. Section \ref{sec,noiseless} considers the noise free (deterministic) version of the problem with average and worst case errors. Approximate (partial) reconstruction is investigated in Section
\ref{sec,distortion}. In Section \ref{sec,noisy} we consider
different noisy models with additive and dilution effects. Finally, we present our conclusions in Section \ref{sec,conclusions}.

\section{Problem setup}
\label{sec,problem setup}
Among a population of $N$ items, $K$ unknown items are of interest. The collection of these $K$ items represents the defective set $\cal G$. The goal is to construct a pooling design, i.e., a collection of tests, to recover the defective set while reducing the number of required tests.
\begin{figure}
\vspace{3mm}
\begin{center}
\includegraphics[width=11.5cm]{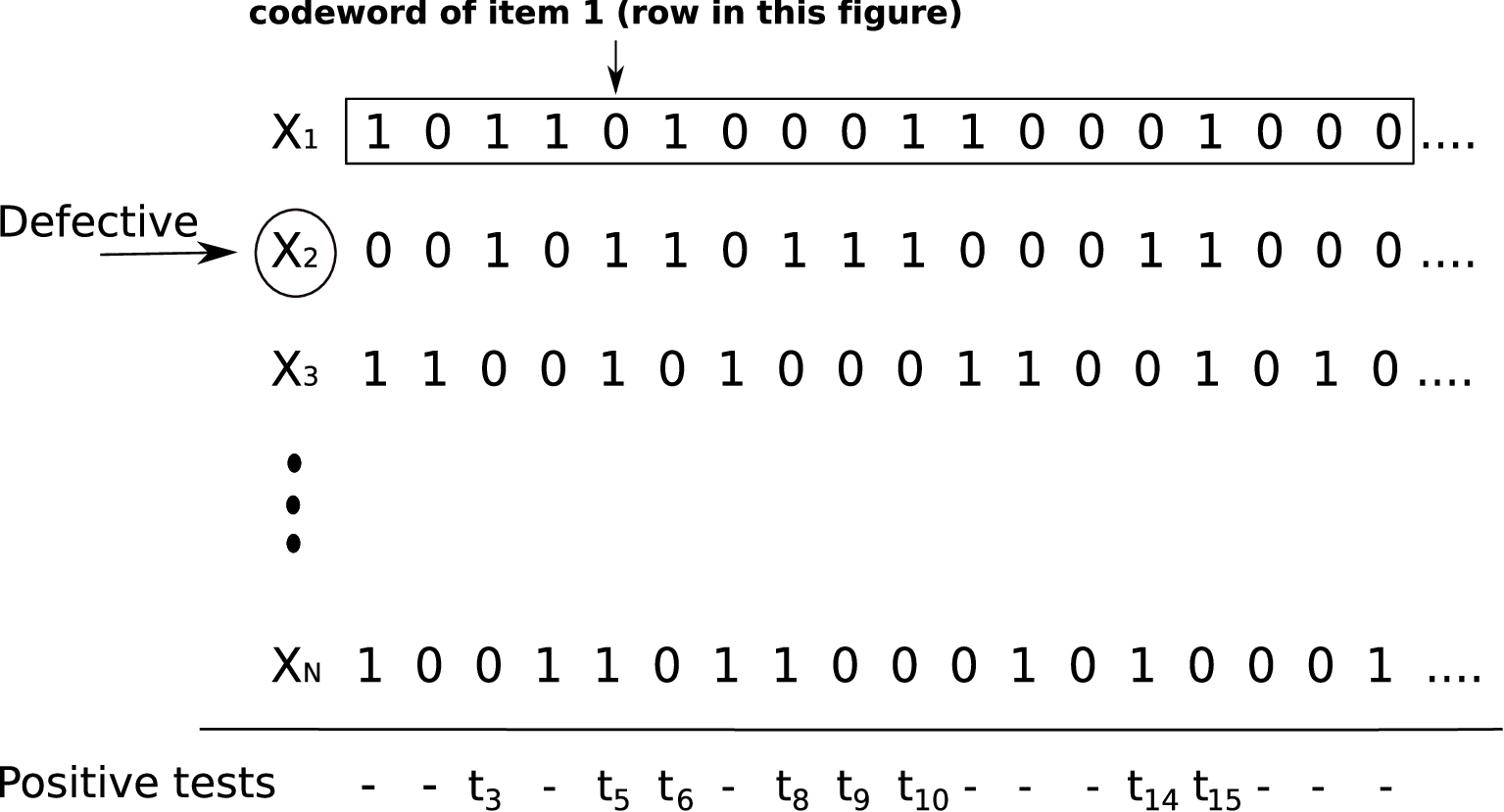}
\caption{A binary matrix defines the assignment of items to tests. Rows are codewords for corresponding items, and columns correspond to pools of items (tests). The entry is $1$ if the item is a member of the designated test and $0$ otherwise. In this example the defective set ${\cal G}=\{2\}$, i.e., only the second item is defective. At the bottom of the figure we show the positive tests. Since this figure illustrates a noise-free example, the outcome of a test is positive if and only if the second item is a member of that test. The goal is to recover the defective set $\cal G$ from $T$ test outcomes.}
\label{fig:codes}
\end{center}
\end{figure}
The idea is illustrated in Fig. \ref{fig:codes}. In this example, the defective set ${\cal G}=\{2\}$, i.e., $K=1$, since only the second item is defective. The binary valued matrix shown in Fig. \ref{fig:codes} represents the measurement matrix defining the assignment of items to tests. The entry is $1$ if the item is a member of the designated test and $0$ otherwise. At the bottom of the figure we highlight the positive tests. The outcome of a test is positive if and only if the second item is a member of that test. Observing the output for a number of tests $T$, the goal is to recover the defective set $\cal G$. While in combinatorial group testing the goal is to find the defective set for the worst-case input, probabilistic group testing requires the average error to be small. Both formulations are considered in this paper in subsequent sections. We assume that the item-test assignment is generated randomly. Before we provide our main result we introduce the notation that will be used throughout the paper.
\subsection{Notation}
\begin{figure}
\vspace{3mm}
\begin{center}
\includegraphics[width=9cm]{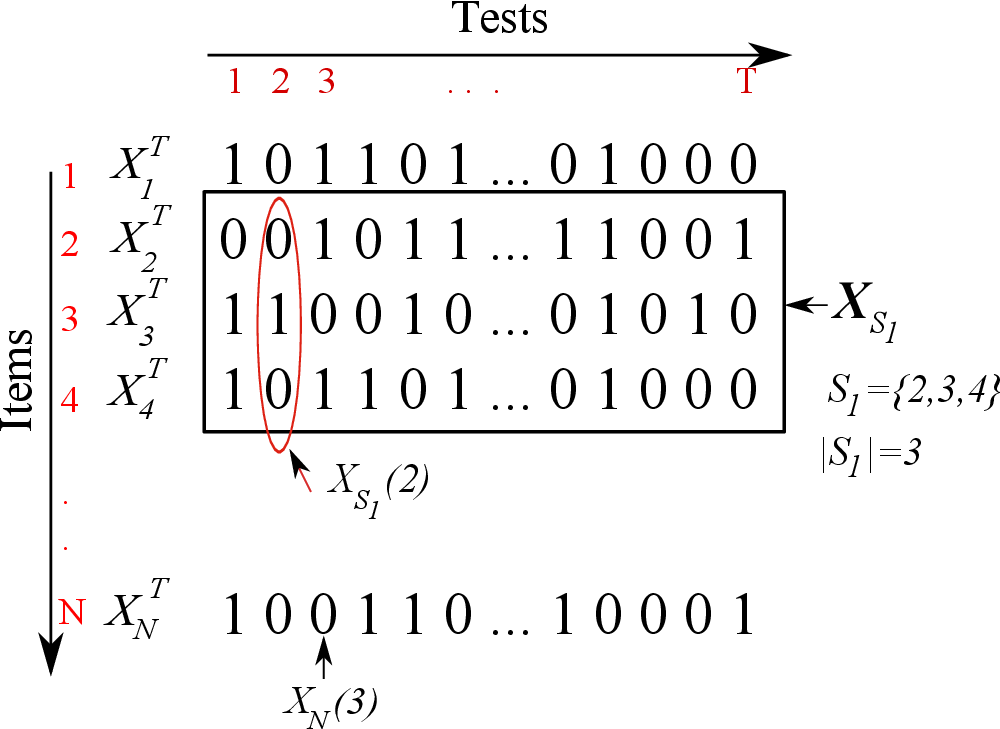}
\caption{The figure illustrates the notation through a simple example. Rows are codewords of length $T$ for corresponding items (denoted $X_j^T$). A defective set $S_1 = \{2,3,4\}$ consists of items $2,3$ and $4$. The corresponding matrix $\emb{X}_{S_1}$ is formed from rows indexed by $S_1$, i.e., consists of rows $2,3,4$. The figure also illustrates the notation $X_{S_1}(2)$ which refers to the vector indexed by the set $S_1$ for test $t=2$. $X_N(3)$ refers to the $3$-rd test for item $N$. Since the matrix is generated i.i.d., the vector $X_{S_1}$ (without a test index) refers to a vector whose components are restricted to the set of components indexed by $S_1$.}
\label{fig:notation}
\end{center}
\end{figure}

We use bold-face to denote matrices, while regular font is used to denote vectors, scalars and entries of matrices. For clarity, Fig.~\ref{fig:notation} illustrates our notation by means of a simple example.
\begin{itemize}
\item $N$ is the total number of items, $K$ is the known number of defectives (or positive items), $p$ denotes the probability that an item is part of a given test, and $T$ is the total number of tests.
\item Codewords: For the $j$-th item, $X_j^T$ is a binary row vector $\in\{0,1\}^T$, with the $t$-th entry $X_j(t)=1$ if the $j$-th item is pooled in test $t$, and $0$ otherwise. Following an information theoretic convention, we call it the $j$-th codeword. 
 The observation vector $Y^T$ is a binary vector of length $T$, with entries equal to $1$ for the tests with positive outcome. Similarly, $Y(t)$ denotes the $t$-th component of the vector $Y^T$.
\item $\emb{X}$: The $N\times T$ measurement matrix, or the codebook, is a collection of $N$ codewords defining the pool design, i.e., the assignment of items to tests. Note that
     \[
    \emb{X}=[X_1^T;X_2^T;\ldots;X_N^T]
     \]
     where each entry represents a row of the matrix $\emb{X}$.
\item Given a subset $S\subset\{1,2 \ldots N\}$ with cardinality $|S|$, the matrix $\emb{X}_S$ is an $|S|\times T$ matrix formed from the rows indexed by $S$. In other words, $\emb{X}_S$ denotes the collection of codewords (each of length $T$) corresponding to the items in $S$. Similarly, $X_S$ denotes a vector, whose components are restricted to the set of components indexed by $S$. Thus, $X_S$ is a column of the matrix $\emb{X}_S$. When indexing by test is needed, $X_S(t)$ is used to specifically denote the $t$-th column of the matrix $\emb{X}_S$, and $X_j(t)$ is the $t$-th component of the vector $X_j^T$.
\item Index the different sets of items of size $K$ as $S_{\omega}$ with index $\omega$. Since there are $N$ items in total, there are $\binom{N}{K}$ such sets, hence
    \begin{equation}
    \omega\in{\cal I}=\left\{1,2,\ldots\binom{N}{K}\right\}
    \end{equation}
    Note that $S_{\omega}$ is a set of $K$ indices corresponding to the $\omega$-th set of items. The empty set is denoted $\emptyset$ and ``$\bigvee$" is used to denote the Boolean OR operation.
\item Hence $\emb{X}_{S_\omega}$ is the matrix of codewords corresponding to the items in the index set $S_{\omega}$.
\end{itemize}

\subsection{Noise-free case}
For the noise-free case, the outcome of the tests $Y^T$ is deterministic. It is the Boolean sum of the codewords corresponding to the defective set $\cal G$. In other words
\begin{equation}
Y^T=\bigvee_{i\in{\cal G}}X_i^T
\label{eq,noise_free}.
\end{equation}

Alternatively, if $R_i \in\{0,1\}$ is an indicator function for the $i$-th item determining whether it belongs to the defective set, i.e., $R_i=1$ if $i\in{\cal G}$ and $R_i=0$ otherwise, then the outcome $Y(t)$ of the $t$-th test in the noise-free case can be written as
\begin{equation}
Y(t)=\bigvee_{i=1}^N X_i(t)R_i
\label{eq:noise_free}
\end{equation}
where $X_i(t)$ is the $t$-th entry of the vector $X_i^T$, or equivalently, the binary entry at cell $(i,t)$ of the measurement matrix $\emb{X}$.

\subsection{Noisy cases}
In this paper we also consider two noisy models, the additive model and the dilution model. However, we point out that our main achievability result is general and is not restricted to these specific noise models or to Boolean channels.
\begin{itemize}
\item Additive Model:
In this model we account for false alarms in the outcome of pooling tests. The outcome of a test can still be $1$ even if no positive items are pooled in that test. This effect is captured by adding independent Bernoulli$(q)$ random variables $W(t)$ to the outcome of the $t$-th test of the noise-free model in Eq.(\ref{eq:noise_free}), i.e.,
\begin{equation}
Y(t)=\left(\bigvee_{i=1}^N X_i(t)R_i\right)\vee W(t)
\label{eq,additive_noise}
\end{equation}
where $W(t) \sim \mbox{Bernoulli}(q), t=1\ldots T$.

\item Dilution Model:
The dilution model refers to the case when a defective item in a pool gets diluted. If all positive items in a given test appear as absent, that could potentially lead to erroneously zero outcomes. This model is motivated by blood dilution due to pooling with other negative tests or imperfectly diluted blood samples. This effect is captured by the Z-channel model of Fig.\ref{fig,z_channel}. A Z-channel with crossover probability $u$ is a binary-input-binary-output channel that flips the input bit $1$ with probability $u$, but maps input bit $0$ to $0$ with probability $1$ (for different Z-channel models the roles of $0$ and $1$ may be interchanged)~\cite{gallager}. In the dilution context, the input $1$ corresponds to the inclusion of a positive item in a test while input $0$ corresponds to the item being absent in that test. The asymmetric crossover captures the dilution effect, that is, in this model we might have misses but no false alarms. The outcome of the $t$-th test can be written as:
\begin{eqnarray}
Y(t)=\bigvee_{i=1}^N {\cal Z}(X_i(t)R_i) \label{eq:dilution_model}
\end{eqnarray}
\end{itemize}
where $\cal Z$ represents the Z-channel model of Fig. \ref{fig,z_channel}. Note that each term on the right hand side is crossed over independently according to this Z-channel model.

\begin{figure}[htbp]
\vspace{3mm}
\centering
\includegraphics[width=5cm]{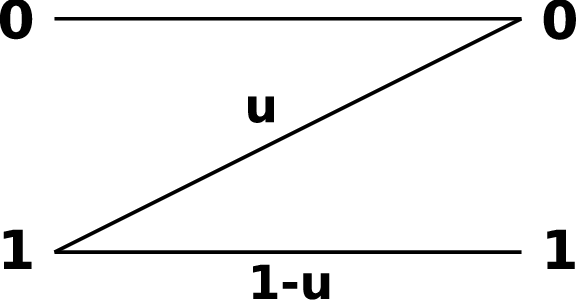}
\caption{Dilution channel (Z-channel): positive items taking part in a given test might probabilistically behave as absent (diluted). In other words, even though a positive item is contained in a given pool, the test's outcome could be negative if the item's presence gets diluted for that specific test. For example, in blood testing the positive sample might get diluted in one or more tests leading to potential misses of infected blood samples.}
\label{fig,z_channel}
\end{figure}
Note that in the additive model, the outcome of testing a pool with no defective items might be erroneously positive, i.e., false positives would occur. On the other hand, the membership of a defective item in a given test might go unnoticed in the dilution model. If all defectives appearing in a given pool are diluted, a false negative occurs. The effect of dilution on the increase in number of tests is expected to be more severe than the effect of additive noise. This is explained by the fact that tests with negative outcomes are generally more informative; while a truly positive test merely indicates that at least one defective is present in the pool, a truly negative test exonerates all the members pooled in that test. With additive noise, a test with a negative outcome is never erroneous. In contrast, dilution diminishes our confidence in pools with negative outcomes, since a seemingly negative outcome would not necessarily mean all pooled members are perfect. Intuitively speaking, additive noise can be potentially mitigated by repetition of tests whereas the dilution effect is more intricate to resolve. This intuition is verified by the results we obtained through theoretical analysis as will be shown in the next sections.

\subsection{Performance criteria}
The different sets of items of size $K$ are indexed $S_{\omega}$, where $\omega\in{\cal I}=\{1,2,\ldots\binom{N}{K}\}$. Define a decoding function $g(.): {\cal Y}^T\rightarrow {\cal I}$. The function $g(.)$ maps the outcome of the $T$ tests, $Y^T\in{\cal Y}^T$, to an index $\omega\in{\cal I}$ corresponding to a specific set of defectives $S_{\omega}$. Now define the conditional error probability $\lambda_{\omega}$ as
\begin{equation}
\lambda_{\omega}=\Pr[g(Y^T ) \ne \omega|{\cal G} = S_\omega,\emb{X}_{S_{\omega}}]=\sum_{Y^T}p_{\omega}(Y^T|\emb{X}_{S_{\omega}})\indic{g(Y^T)\ne \omega}
\end{equation}
where $\indic{.}$ is an indicator function which takes the value $1$ when its argument is realized and $0$ otherwise, i.e., it takes the value $1$ when the defective set is misclassified. The conditional probability $p_{\omega}(Y^T|\emb{X}_{S_{\omega}})$ is the probability that $Y^T$ is the outcome of the $T$ tests given that $S_{\omega}$ is the defective set with corresponding codewords matrix $\emb{X}_{S_{\omega}}$. The conditional probability defines a group testing channel which is analogous to a communication channel where the output sequence depends on the transmitted input sequence, the message being the defective set and the encoded sequence being the matrix of codewords corresponding to that set. The probability $\lambda_{\omega}$ is the probability of error conditioned on a given defective set $\omega$ and the codeword set, i.e., the probability that the decoded set is not the true defective set given that $S_{\omega}$ is the true defective set with codeword set $\emb{X}_{S_{\omega}}$. Note that for the deterministic noise-free case this simplifies to $\lambda_{\omega}=\indic{g(Y^T)\ne \omega}$. Averaging over all possible inputs $\omega$, we define the average error probability $\lambda$ as:
\begin{equation}
\lambda=\frac{1}{\binom{N}{K}}\sum_{\omega} \lambda_{\omega}
\end{equation}

For the aforementioned models (noiseless and noisy), we prove achievable and converse bounds on the total number of tests $T$ as $N,K\rightarrow\infty$. Namely, we consider the following criteria:
\begin{itemize}
\item Arbitrarily small average error probability\footnote{We follow the standard reasoning approach in \cite{coverbook,gallager} where we consider the error probability averaged all codebook realizations. An arbitrarily small average error probability in turn implies the existence of at least one good codebook.}:
\[
\lambda=\frac{1}{\binom{N}{K}}\sum_{\omega} \lambda_{\omega}
\]
\item Worst-case error probability:
\[
\lambda_{\max}=\max_{\omega} \lambda_{\omega}
\]
\item Partial reconstruction: In this case, we are satisfied with approximate reconstruction of the defective set. Let $d$ be a distance function between the decoded set $g(Y^T)$ and the index $\omega$ of the true defective set such that $d(g(Y^T),\omega)$ is equal to the number of misses. Hence, if $g(Y^T)=\hat{\omega}$, then
    \[
    d(g(Y^T),\omega) = |S_{\omega}\backslash S_{\hat{\omega}}|
    \]
    is the cardinality of the set of missed items. Given $K$ declared candidates, an error occurs only if the number of missed items is greater than $\alpha K$, i.e.,
\[
\lambda_{\omega}=\sum_{Y^T}p_{\omega}(Y^T|\emb{X}_{S_{\omega}})\indic{d(g(Y^T),\omega)>\alpha K}
\]
\end{itemize}
In the following section we will derive our main result. We will
prove a sufficient condition on the number of tests $T$. The result is general as it applies to the noise-free and the noisy
versions of the problem as we elaborate in the following sections.

\section{Main Result: Achievable bound}
\label{sec,main_result} To derive an achievable bound on the number
of tests $T$, we show how the group testing problem can be mapped to
an equivalent channel model. Using random coding and maximum likelihood
decoding we upper bound the error probability, i.e., the probability of misclassifying the defective set.

\subsection{Random matrix generation and the encoding process}
The binary measurement matrix is randomly generated. Associated with each item is a codeword that represents its assignment to tests. Assume that the codewords are generated randomly and independently according to some distribution $q_T$. The probability of a particular codebook $\emb{X}$ assigning one codeword to each of the $N$ items, is $Q(\emb{X})=\prod_{j=1}^N q_T(X_j^T)$, where $q_T$ is a probability assignment on the set of input sequences of length $T$. Since we will be assuming independence across tests and across items, to simplify notation we will use $Q(\emb{F})$ to denote the distribution of a matrix $\emb{F}$ of arbitrary size such that
\[
Q(\emb{F}) = \prod_{i=1}^{n_1}\prod_{t=1}^{n_2} q(F_i(t))
\]
where $F$ is an $n_1\times n_2$ matrix with $n_1\in\{1,\ldots N\}$ and $n_2\in\{1,\ldots T\}$ and $q(F_i(t))$ denotes the distribution of the $(i,t)$-th entry.

\begin{figure}
\centering
\includegraphics[width=7cm]{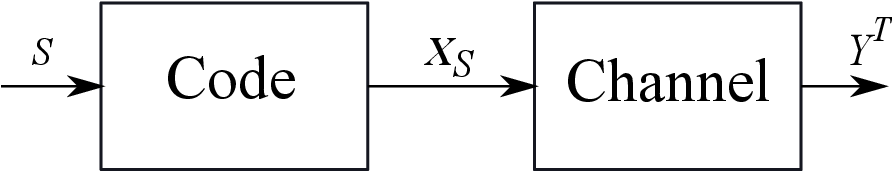}
\caption{A set $S$ is the message that gets mapped to a code $\emb{X}_S$. The encoded message is transmitted through the group testing channel and produces the output $Y^T$.}
\label{fig:channel_model}
\end{figure}

A defective set $S_{\omega}$ corresponds to a collection of $K$ codewords, which is a $K\times T$ matrix $\emb{X}_{S_{\omega}}$, and can be thought of as an encoded message transmitted through the group testing channel. In other words, the encoder $f:{\cal I}\rightarrow\{0,1\}^{K\times T}$, maps an index $\omega\in{\cal I}$ (defective set $S_{\omega}$) to a matrix of codewords $\emb{X}_{S_{\omega}}\in\{0,1\}^{K\times T}$, where $\{0,1\}^{K\times T}$ is the space of binary $K\times T$ matrices. The encoded message $f(\omega)=\emb{X}_{S_{\omega}}$ is then transmitted through a channel with transition probability $p_{\omega}(Y^T|\emb{X}_{S_{\omega}})$ conditioned on the event ${\cal G}=S_{\omega}$. The channel model is illustrated in Fig.\ref{fig:channel_model}.

Since the channel is memoryless, i.e., each test outcome $Y(t)$ depends only on the corresponding input $X_{S_{\omega}}(t)$, the probability of the output sequence $Y^T=(Y(1), Y(2), \ldots, Y(T))$ given the input $\emb{X}_{S_{\omega}} = (X_{S_{\omega}}(1), X_{S_{\omega}}(2), \ldots, X_{S_{\omega}}(T))$ is given by
\[
p_{\omega}(Y^T|\emb{X}_{S_{\omega}}) = \prod_{t=1}^T p_{\omega}(Y(t)|X_{S_{\omega}}(t)).
\]
Conditioned on the event ${\cal G}=S_{\omega}$ the $K$ codewords and the observed $Y^T$ have a joint distribution
\[
p_{\omega}(\emb{X}_{S_{\omega}},Y^{T})=\prod_{t=1}^{T}Q(X_{S_{\omega}}(t))p_{\omega}(Y(t)|X_{S_{\omega}}(t))
\]
Unless needed for clarity, we will often drop the suffix $\omega$ in $p_{\omega}(Y^{T}|\emb{X}_{S_{\omega}})$ to simplify notation since the conditioning index will be generally clear.
%
\bigbreak
\subsection{Decoder}
Decoding is achieved using ML decoding ~\cite{gallager}. The decoder goes through all $\binom{N}{K}$ possible sets of size $K$, where $K$ is the size of the defective set, and chooses the set that is most likely. The decoding rule is thus defined by: given the tests'outcomes $Y^T$, choose $\omega^*$ for which
\begin{equation}
p(Y^T|\emb{X}_{S_{\omega^*}})> p(Y^T|\emb{X}_{S_\omega});~~~~\forall \omega\ne \omega^*
\end{equation}
i.e., choose the set for which the given $Y^T$ is most likely\footnote{When two or more sets are in tie, the decoder can choose any of those sets. This is accounted for in the probability of error analysis.} given $\omega$. An error occurs if any set other than the true defective set is more likely. This ML decoder minimizes the error probability assuming uniform prior on the input messages (defective sets). Next, we derive an upper bound on the average error probability of the ML decoder, where the average is taken over defective sets and ensembles of codewords.

\subsection{Probability of error analysis}
Given the random codebook generation, let $P_e$ denote the average
probability of error, averaged over all codebooks $\emb{X}$, and over all sets
of size $K$, i.e.,
\begin{eqnarray}
P_e&=&\sum_{\emb{X}}Q({\emb{X}})\lambda({\emb{X}})\nonumber\\
&=&\sum_{\emb{X}}Q({\emb{X}})\frac{1}{\binom{N}{K}}\sum_{\omega} \lambda_{\omega}({\emb{X}})\nonumber\\
&=&\frac{1}{\binom{N}{K}}\sum_{\omega} \sum_{\emb{X}}Q({\emb{X}})\lambda_{\omega}({\emb{X}})
\end{eqnarray}

By symmetry of the codebook construction, $\sum_{\emb{X}}Q({\emb{X}})\lambda_{\omega}({\emb{X}})$ does not depend on $\omega$ (and consequently not on the set $S_{\omega}$). Thus,
\begin{equation}
P_e=\sum_{\emb{X}}Q({\emb{X}})\lambda_{\omega}({\emb{X}})=P_{e|\omega}
\end{equation}
where $P_{e|\omega}$ is the error probability conditioned on $\omega$ averaged over all codebooks. In other words, the average error probability does not depend on the input $\omega$ due to averaging over randomly generated codebooks. Hence, if $\omega_0$ is the index of the true defective set we can assume without loss of generality that $\omega_0 = 1$, i.e., $S_1$ is the true defective set.

To simplify the exposition we introduce some further notation. As pointed out earlier, the matrix $\emb{X}_S$ is formed from rows indexed by the set $S$. For any $2$ sets $S_i$ and $S_j$, we define $S_{i,j}$, $S_{i^c,j}$, and $S_{i,j^c}$ as the overlap set, the set of indices in $S_j$ but not in $S_i$, and the set of indices in $S_i$ but not in $S_j$, respectively. Namely,
\begin{align}
S_{i,j} &= S_i\cap S_j \mbox{~~~overlap}\nonumber\\
S_{i^c,j} &= S_i^c\cap S_j \mbox{~~~in $j$ but not in $i$}\nonumber\\
S_{i,j^c} &= S_i\cap S_j^c \mbox{~~~in $i$ but not in $j$}\nonumber
\end{align}

Now define the error event $E_i$ as the event that a set which differs from the defective set $S_1$ in exactly $i$ items is selected by the decoder. The probability of such an event is denoted $P(E_i)$. The event $E_i$ implies that there exists some set which differs from the defective set in $i$ items and is more likely. Hence,
\begin{align}
P(E_i) \leq \Pr\Big[&\exists j\ne 1: p_j(Y^T|\emb{X}_{S_j})\geq p_1(Y^T|\emb{X}_{S_1})\nonumber\\
&\mbox{where}~~|S_{1^c,j}|=|S_{1,j^c}|=i,~\mbox{and}~~|S_1|=|S_j|=K\Big]
\end{align}

The probability $P(E_i)$ can be written as a summation over all inputs $\emb{X}_{S_1}$ and all test outcomes $Y^T$
\begin{equation}
P(E_i)= \sum_{\emb{X}_{S_1}}\sum_{Y^T} Q(\emb{X}_{S_1})p_1(Y^T|\emb{X}_{S_1})\Pr[E_i| \omega_0 = 1, \emb{X}_{S_1}, Y^T]
\label{eq:P(Ei)}
\end{equation}
where $\Pr[E_i |\omega_0 = 1, \emb{X}_{S_1}, Y^T]$ is the probability of decoding error in exactly $i$ items, conditioned on message $\omega_0=1$,
the selection of a particular $\emb{X}_{S_1}$ as the codewords for the set $S_1$, and on the reception of a sequence $Y^T$.

Using the union bound, the conditional error probability averaged over ensembles of codewords is upper bounded by,
\begin{align}
P_{e|1}&\leq\sum_{i=1}^K P(E_i)\nonumber\\
&=\sum_{i=1}^K \sum_{\emb{X}_{S_1}}\sum_{Y^T} Q(\emb{X}_{S_1})p(Y^T|\emb{X}_{S_1})\Pr[E_i |\omega_0=1, \emb{X}_{S_1}, Y^T]
\label{eq:P_e,1}
\end{align}

Next we state our main result. We need to introduce new notation to describe the result. Define $\Xi_{S}^{\{i\}}$ as the set of tuples $({\cal S}^1,{\cal S}^2)$ partitioning the defective set $S$ into disjoint sets ${\cal S}^1$ and ${\cal S}^2$ with cardinalities $i$ and $K-i$, respectively, i.e.,
\begin{align}
\Xi_S^{\{i\}} = \left\{({\cal S}^1,{\cal S}^2): {\cal S}^1\bigcap{\cal S}^2=\emptyset, {\cal S}^1\bigcup{\cal S}^2=S, |{\cal S}^1|=i, |{\cal S}^2|=K-i  \right\}
\label{eq:Xi}
\end{align}
We let $I(X_{{\cal S}^1};X_{{\cal S}^2},Y)$ denote the mutual information \cite{coverbook} between $X_{{\cal S}^1}$ and $(X_{{\cal S}^2},Y)$ defined as,
\begin{align}
I(X_{{\cal S}^1};X_{{\cal S}^2},Y)=\sum_Y\sum_{X_{{\cal S}^2}}\sum_{X_{{\cal S}^1}} Q(X_{{\cal S}^1})p(Y,X_{{\cal S}^2}|X_{{\cal S}^1})\log\frac{p(Y,X_{{\cal S}^2}|X_{{\cal S}^1})}{\sum_{X_{{\cal S}^1}} Q(X_{{\cal S}^1})p(Y,X_{{\cal S}^2}|X_{{\cal S}^1})}\,\,.\nonumber
\end{align}
The following theorem provides a sufficient condition on the number of tests $T$ for an arbitrarily small average error probability.
\begin{theorem}
(Sufficiency).
Let $N$ be the size of a population of items with the defective set $S$ of cardinality $K$. If the number tests $T$ is such that
\begin{eqnarray} \label{eq:pos_error_exp}
T>(1+\epsilon)\cdot\max_{i:({\cal S}^1,{\cal S}^2)\in\Xi_S^{\{i\}}} \frac{\log \binom{N-K}{i}\binom{K}{i}}{I(X_{{\cal S}^1};X_{{\cal S}^2},Y)}, ~~i=1, 2, \ldots K
\end{eqnarray}
then asymptotically the average error probability approaches zero, namely,
\begin{align}
\lim_{K\rightarrow\infty}\lim_{N\rightarrow\infty}P_e\rightarrow 0
\end{align}
where $\epsilon>0$ is a constant independent of $N$ and $K$.
\label{thm:main_result}
\end{theorem}
\begin{remark} Theorem~\ref{thm:main_result} provides a sufficient condition for the case where \emph{$N$ scales to infinity for every fixed K}. The question arises as to what happens when both $N$ and $K$ scale at the same rate. We will describe this case in Section~\ref{sec:general_case}.
\end{remark}

Theorem~\ref{thm:main_result} follows from a tight bound---based on characterization of error exponents as in \cite{gallager}---on the error probability $P(E_i)$. We will show that the error exponent, $E_o(\rho)$, is described by:
\begin{align}
E_o(\rho) = -\log\sum_{Y\in\{0,1\}}\sum_{X_{{\cal S}^2}}\left[\sum_{X_{{\cal S}^1}}Q(X_{{\cal S}^1}) p(Y,X_{{\cal S}^2}|X_{{\cal S}^1})^{\frac{1}{1+\rho}}\right]^{1+\rho} ~~~0\leq\rho\leq 1
\label{eq:E0}
\end{align}
where, $({\cal S}^1,{\cal S}^2)\in\Xi_{S}^{\{i\}}$, defined in (\ref{eq:Xi}), denote any disjoint partitions of the defective set $S_1$ with cardinalities $i$ and $K-i$, respectively. $X_{{\cal S}^1}$ and $X_{{\cal S}^2}$ are the corresponding disjoint partitions of the $K\times 1$ input $X_{S_1}$ of lengths $i\times 1$ and $(K-i)\times 1$, respectively. We then have the following result:
\begin{lemma}
\label{thm:P(Ei)}
The probability of the error event $E_i$ defined in Eq. \ref{eq:P(Ei)} that a set which differs from the defective set $S_1$ in exactly $i$ items is selected by the ML decoder (averaged over all codebooks and test outcomes) is bounded from above by
\begin{align}
P(E_i)\leq 2^{-T\left(E_o(\rho)-\rho\frac{\log\binom{N-K}{i}\binom{K}{i}}{T}\right)}
\label{eq:error_exp}
\end{align}
\end{lemma}
We are now ready to prove Theorem~\ref{thm:main_result}.

\subsection{Proof of Theorem \ref{thm:main_result}}
Now we can readily prove our main result. First, we need to derive a sufficient condition for the error exponent of the error probability $P(E_i)$ in (\ref{eq:error_exp}) to be positive and to drive the error probability to zero as $N\rightarrow\infty$. Specifically,
\begin{align}
Tf(\rho)= TE_o(\rho) - \rho\log\binom{N-K}{i}\binom{K}{i} \rightarrow\infty
\label{eq:f_rho_condition}
\end{align}
where
\[
f(\rho) = E_o(\rho)-\rho\frac{\log\binom{N-K}{i}\binom{K}{i}}{T}
\]
and where $E_o(\rho)$ is defined in (\ref{eq:E0}). 

To establish Eq.~\ref{eq:pos_error_exp} we follow the argument in \cite{gallager}. Note that $f(0)=0$. Since the function $f(\rho)$ is differentiable and has a power series expansion, for a sufficiently small $\delta$, we get by Taylor series expansion in the neighborhood of $\rho\in[ 0,\delta]$ that,
\[
f(\rho) = f(0) + \rho\frac{df}{d\rho}\Big\vert_{\rho=0} + O(\rho^2)
\]
But we can show that
\begin{align}
\frac{\partial E_o}{\partial \rho}\Big\vert_{\rho=0} = \sum_Y\sum_{X_{{\cal S}^2}}\Bigg[&\sum_{X_{{\cal S}^1}} Q(X_{{\cal S}^1})p(Y,X_{{\cal S}^2}|X_{{\cal S}^1})\log p(Y,X_{{\cal S}^2}|X_{{\cal S}^1})\nonumber\\
-&\sum_{X_{{\cal S}^1}} Q(X_{{\cal S}^1})p(Y,X_{{\cal S}^2}|X_{{\cal S}^1})\log\sum_{X_{{\cal S}^1}} Q(X_{{\cal S}^1})p(Y,X_{{\cal S}^2}|X_{{\cal S}^1})\Bigg]
\end{align}
which simplifies to
\begin{align}
\frac{\partial E_o}{\partial \rho}\Big\vert_{\rho=0}  &=  \sum_Y\sum_{X_{{\cal S}^2}}\sum_{X_{{\cal S}^1}} Q(X_{{\cal S}^1})p(Y,X_{{\cal S}^2}|X_{{\cal S}^1})\log\frac{p(Y,X_{{\cal S}^2}|X_{{\cal S}^1})}{\sum_{X_{{\cal S}^1}} Q(X_{{\cal S}^1})p(Y,X_{{\cal S}^2}|X_{{\cal S}^1})}\nonumber\\
& = I(X_{{\cal S}^1};X_{{\cal S}^2},Y)
\end{align}
Now it is easy to see that with $(1+\epsilon)\frac{\log\binom{N-K}{i}\binom{K}{i}}{T}<I(X_{{\cal S}^1};X_{{\cal S}^2},Y)$ for some constant $\epsilon>0$, the condition in (\ref{eq:f_rho_condition}) is satisfied, i.e., $Tf(\rho)\rightarrow\infty$ as $N\rightarrow\infty$. To argue this we first note that from the Lagrange form of the Taylor Series expansion (an application of the mean value theorem) we can write $E_o(\rho)$ in terms of its first derivative evaluated at zero and a remainder term, i.e.,
\begin{align}
E_o(\rho) = E_o(0) + \rho E_o'(0) + {\rho^2 \over 2} E_o''(\psi)
\label{eq:lagrange}
\end{align}
for some $\psi \in [0,\rho]$. Hence, for the choice of $T$ in (\ref{eq:pos_error_exp}) we have
\begin{align*}
Tf(\rho)\geq T\left(\rho\frac{\epsilon}{1+\epsilon}I(X_{{\cal S}^1};X_{{\cal S}^2},Y)-\rho^2 C I(X_{{\cal S}^1};X_{{\cal S}^2},Y)\right)
\end{align*}
where $C=-\frac{|E_o''(\psi)|}{I(X_{{\cal S}^1};X_{{\cal S}^2},Y)}$ which might depend on $K$. Note that, $C$ being a continuous function is bounded on the closed unit interval. Now if we choose $\rho\leq\frac{\epsilon'}{C}$, where $\epsilon' = \frac{\epsilon}{1+\epsilon}$, then $f(\rho)=\delta$ for some $\delta>0$ which does not depend on $N$ or $T$. It follows that $Tf(\rho)\rightarrow\infty$ as $N\rightarrow\infty$.

We have just shown that for fixed $K$, $T > (1+\epsilon)\cdot\frac{\log\binom{N-K}{i}\binom{K}{i}}{I(X_{{\cal S}^1};X_{{\cal S}^2},Y)}$ is sufficient to ensure an arbitrarily small $P(E_i)$. Since the average error probability $P_e\leq\sum_{i=1}^K P(E_i)$, it follows that for any fixed $K$, $\lim_{N\rightarrow\infty} \sum_{i=1}^K P(E_i)=0$. Consequently, since this is true for any $K$, $\lim_{K\rightarrow\infty}\lim_{N\rightarrow\infty}\sum_{i=1}^K P(E_i)=0$.
Theorem \ref{thm:main_result} now follows.

\subsection{Proof of Lemma \ref{thm:P(Ei)}} \label{subsec.mainresult}
To build intuition we will first show the following weaker bound:
\begin{align}
P(E_i)\leq 2^{-T\left(E_o(\rho)-\frac{\log\binom{N-K}{i}\binom{K}{i}}{T}\right)}
\label{eq:error_exp_weak}
\end{align}
Note that the main difference between the above equation and Lemma~\ref{thm:P(Ei)} is the missing $\rho$ term multiplying the binomial expression. The main result follows along the same lines and is described in detail in the appendix. It is based on exploiting two types of symmetry: the symmetry of codebook construction; and the symmetry of the channel output to permutations of the rows of matrix $\emb{X}$. Note that it is possible to obtain qualitatively similar results as in Table~\ref{tab:summary} using this weaker bound. However, the expression in Theorem~\ref{thm:main_result} does not turn out to be as explicit with this bound.

To prove this weaker result we denote by ${\cal A}$
\begin{align}
{\cal A}=\{\omega\in{\cal I}: |S_{1^c,\omega}|=i, |S_{\omega}|=K\}
\label{eq:setA}
\end{align}
the set of indices corresponding to sets of $K$ items that differ from the true defective set $S_1$ in exactly $i$ items. We can establish that,
\begin{align} \label{eq.ratio}
\Pr[E_i | \omega_0 = 1, \emb{X}_{S_1}, Y^T] &\leq
\sum_{\omega \in {\cal A}}\sum_{\substack{\emb{X}_{S_{1^c,\omega}}}} Q(\emb{X}_{S_{1^c,\omega}})\frac{p_{\omega}(Y^T,\emb{X}_{S_{1,\omega}}|\emb{X}_{S_{1^c,\omega}})^s}{p_1(Y^T,\emb{X}_{S_{1,\omega}}|\emb{X}_{S_{1,\omega^c}})^s}\\ \nonumber
& = \sum_{S_{1,\omega}} \sum_{S_{1^c,\omega}}\sum_{\substack{\emb{X}_{S_{1^c,\omega}}}} Q(\emb{X}_{S_{1^c,\omega}})\frac{p_{\omega}(Y^T,\emb{X}_{S_{1,\omega}}|\emb{X}_{S_{1^c,\omega}})^s}{p_1(Y^T,\emb{X}_{S_{1,\omega}}|\emb{X}_{S_{1,\omega^c}})^s}
\end{align}
Inequality (\ref{eq.ratio}) is established in the Appendix. It follows that,
\begin{align*}
\Pr[E_i | \omega_0 = 1, \emb{X}_{S_1}, Y^T] &\stackrel{(a)}{\leq}
\left (\sum_{S_{1,\omega}} \sum_{S_{1^c,\omega}} \sum_{\substack{\emb{X}_{S_{1^c,\omega}}}} Q(\emb{X}_{S_{1^c,\omega}})\frac{p_{\omega}(Y^T,\emb{X}_{S_{1,\omega}}|\emb{X}_{S_{1^c,\omega}})^s}{p_1(Y^T,\emb{X}_{S_{1,\omega}}|\emb{X}_{S_{1,\omega^c}})^s} \right )^{\rho}\\
\nonumber &\stackrel{(b)}{\leq}
\left (\sum_{S_{1,\omega}}\binom{N-K}{i} \sum_{\substack{\emb{X}_{S_{1^c,\omega}}}} Q(\emb{X}_{S_{1^c,\omega}})\frac{p_{\omega}(Y^T,\emb{X}_{S_{1,\omega}}|\emb{X}_{S_{1^c,\omega}})^s}{p_1(Y^T,\emb{X}_{S_{1,\omega}}|\emb{X}_{S_{1,\omega^c}})^s} \right )^{\rho}
\\ \nonumber &\stackrel{(c)}{\leq}
\binom{N-K}{i}\sum_{S_{1,\omega}}\left (\sum_{\substack{\emb{X}_{S_{1^c,\omega}}}} Q(\emb{X}_{S_{1^c,\omega}})\frac{p_{\omega}(Y^T,\emb{X}_{S_{1,\omega}}|\emb{X}_{S_{1^c,\omega}})^s}{p_1(Y^T,\emb{X}_{S_{1,\omega}}|\emb{X}_{S_{1,\omega^c}})^s} \right )^{\rho}
~~ \forall s>0,\,\, 0\leq \rho \leq 1
\end{align*}
Inequality (a) follows from the fact that $\Pr[E_i | \omega_0 = 1, \emb{X}_{S_1}, Y^T] \leq 1$. Consequently, if $U$ is an upperbound of this probability then it follows that, $\Pr[E_i | \omega_0 = 1, \emb{X}_{S_1}, Y^T] \leq U^{\rho}$ for $\rho \in [0,1]$. Inequality (b) follows from symmetry of codebook construction, namely, the inner summation is only dependent on the values of $\emb{X}_{S_{1^c,\omega}}$ and not on the items in the set $S_{1^c,\omega}$. There are exactly $\binom{N-K}{i}$ possible sets $S_{1^c,\omega}$ hence the binomial expression. Note that the sum over $S_{1,\omega}$ cannot be further simplified. This is due to the fact that $\emb{X}_{S_{1,\omega}}$ is already specified since we have conditioned on $\emb{X}_{S_1}$. Since $\emb{X}_{S_1}$ is fixed, the inner sum need not be equal for all sets $S_{1,\omega},\omega\in{\cal A}$. {\it Indeed, the main difference between the weaker and stronger bound is this issue. In the stronger bound we incorporate the symmetry both in the missed as well as common components.} Finally, (c) follows from standard observation that sum of positive numbers raised to $\rho$-th power for $\rho < 1$ is smaller than the sum of the $\rho$-th power of each number.

We now substitute for the conditional error probability derived above and follow the steps below:
\begin{align} \nonumber
P(E_i)&=\sum_{\emb{X}_{S_1}}\sum_{Y^T} p_1(\emb{X}_{S_1},Y^T)\Pr[E_i | \omega_0 = 1, \emb{X}_{S_1}, Y^T]
\\ \nonumber
& \leq
\binom{N-K}{i} \sum_{S_{1,\omega}} \sum_{Y^T}\sum_{\emb{X}_{S_1}} p_1(\emb{X}_{S_1},Y^T)\left (\sum_{\substack{\emb{X}_{S_{1^c,\omega}}}}
Q(\emb{X}_{S_{1^c,\omega}})\frac{p_{\omega}(Y^T,\emb{X}_{S_{1,\omega}}|\emb{X}_{S_{1^c,\omega}})^s}{p_1(Y^T,\emb{X}_{S_{1,\omega}}|\emb{X}_{S_{1,\omega^c}})^s} \right )^{\rho}
\end{align}
Due to symmetry the summation over sets $S_{1,\omega}$ does not depend on $\omega$. Since there are $\binom{K}{K-i}$ sets $S_{1,\omega}$ we get,
\begin{align*}
P(E_i)& \leq
\binom{N-K}{i} \binom{K}{i} \sum_{Y^T}\sum_{\emb{X}_{S_1}} p_1(\emb{X}_{S_1},Y^T)\left (\sum_{\substack{\emb{X}_{S_{1^c,\omega}}}} Q(\emb{X}_{S_{1^c,\omega}})\frac{p_{\omega}(Y^T,\emb{X}_{S_{1,\omega}}|\emb{X}_{S_{1^c,\omega}})^s}{p_1(Y^T,\emb{X}_{S_{1,\omega}}|\emb{X}_{S_{1,\omega^c}})^s} \right )^{\rho}
\\ \nonumber & \leq
\binom{N-K}{i} \binom{K}{i} \sum_{Y^T}\sum_{\emb{X}_{S_{1,\omega^c}}}\sum_{\emb{X}_{S_{1,\omega}}} Q(\emb{X}_{S_{1,\omega^c}})p_1(\emb{X}_{S_{1,\omega}},Y^T\mid \emb{X}_{S_{1,\omega^c}})\\ \nonumber  & ~~~~~~~~~~~~~~~~~~~~~~~~~~~~~~~~~~~~~~~~~~~~~~~~~~~~~~~~~~~
\left (\sum_{\substack{\emb{X}_{S_{1^c,\omega}}}} Q(\emb{X}_{S_{1^c,\omega}})\frac{p_{\omega}(Y^T,\emb{X}_{S_{1,\omega}}|\emb{X}_{S_{1^c,\omega}})^s}{p_1(Y^T,\emb{X}_{S_{1,\omega}}|\emb{X}_{S_{1,\omega^c}})^s} \right )^{\rho}
\\ \nonumber & =
\binom{N-K}{i} \binom{K}{i} \sum_{Y^T} \sum_{\emb{X}_{S_{1,\omega^c}}}\sum_{\emb{X}_{S_{1,\omega}}} Q(\emb{X}_{S_{1,\omega^c}})p_1^{1-s\rho}(\emb{X}_{S_{1,\omega}},Y^T\mid \emb{X}_{S_{1,\omega^c}}) \\
\nonumber & ~~~~~~~~~~~~~~~~~~~~~~~~~~~~~~~~~~~~~~~~~~~~~~~~~~~~~~~~~~~
\left (\sum_{\substack{\emb{X}_{S_{1^c,\omega}}}} Q(\emb{X}_{S_{1^c,\omega}})p_{\omega}(Y^T,\emb{X}_{S_{1,\omega}}|\emb{X}_{S_{1^c,\omega}})^s \right )^{\rho}
\\ \nonumber & =
\binom{N-K}{i} \binom{K}{i} \sum_{Y^T} \sum_{\emb{X}_{S_{1,\omega}}} \left ( \sum_{\emb{X}_{S_{1,\omega^c}}} Q(\emb{X}_{S_{1,\omega^c}})p_1^{1/(1+\rho)}(\emb{X}_{S_{1,\omega}},Y^T\mid \emb{X}_{S_{1,\omega^c}}) \right )^{1+\rho}
\\ \nonumber & =
\binom{N-K}{i} \binom{K}{i} \sum_{Y} \sum_{X_{S_{1,\omega}}} \left ( \sum_{X_{S_{1,\omega^c}}} Q(X_{S_{1,\omega^c}})p_1^{1/(1+\rho)}(X_{S_{1,\omega}},Y^T\mid X_{S_{1,\omega^c}}) \right )^{T(1+\rho)}
\end{align*}
where the last step follows due to independence of the tests. The step before the last follows by noting that from symmetry $\emb{X}_{S_{1^c,\omega}}$ is just a dummy variable and can be replaced by $\emb{X}_{S_{1,\omega^c}}$. This establishes the weaker bound in (\ref{eq:error_exp_weak}). Further details about the proof of Lemma \ref{thm:P(Ei)} are described in Appendix A.

\subsection{General case where $K=o(N)$}
\label{sec:general_case}
We explicitly differentiate between two different scaling regimes considered in the paper. The result of Theorem \ref{thm:main_result} we just presented provides a sufficient condition for the case where \emph{$N$ scales to infinity for every fixed K} ensuring that
\[
\lim_{K\rightarrow\infty}\lim_{N\rightarrow\infty}P_e\rightarrow 0.
\]
As we mentioned earlier, since the error probability averaged over all codebooks is arbitrarily small, this in turn implies the existence of at least one good codebook and hence $\lambda\rightarrow 0$.
However, we point out that the result in Theorem \ref{thm:main_result} can be extended to the more general case where \emph{both $N$ and $K$ are allowed to scale simultaneously such that $K=o(N)$}. From the Lagrange form of the Taylor Series expansion we can write $E_o(\rho)$ in terms of its first derivative evaluated at zero and a remainder term which depends on the second derivative (\ref{eq:lagrange}).
We have already shown that $E_o(0)=0$ and $E_o'(0) = I(X_{\cs^1}; Y \mid X_{\cs^2})>0$. Consequently, we need to lower bound $E_o(\rho)$ by taking the worst-case second derivative. Establishing this result requires a more careful analysis and hence we present it in Appendix C. It is shown that the same sufficient condition on the number of tests $T$ in Theorem \ref{thm:main_result} holds up to an extra polylog($K$) factor for an arbitrarily small average error probability, namely,
\[
\lim_{\substack{N\rightarrow\infty \\ K=o(N)}} P_e\rightarrow 0
\]

\bigbreak
\noindent{\bf Relation to channel coding:}\\
The mutual information $I(X_{{\cal S}^1};X_{{\cal S}^2},Y)$ between $X_{{\cal S}^1}$ and $(X_{{\cal S}^2},Y)$ is the relative entropy between the joint distribution $p(X_{{\cal S}^1},X_{{\cal S}^2},Y)$ and the product distribution $p(X_{{\cal S}^1})p(X_{{\cal S}^2},Y)$. It is a measure of the amount of information that the variables $X_{{\cal S}^2}$ and $Y$ contain about $X_{{\cal S}^1}$ ~\cite{coverbook}. Intuitively this means that the upper bound on the error probability $P(E_i)$, where the event $E_i$ represents the event that the decoded set is misclassified in exactly $i$ items (replacing $i$ codewords of the true set with $i$ independent codewords), scales exponentially with the negative of the per-test mutual information between $i$ codewords of the set, and the remaining $K-i$ codewords and the output.

\begin{figure}
\centering
\includegraphics[width=7cm]{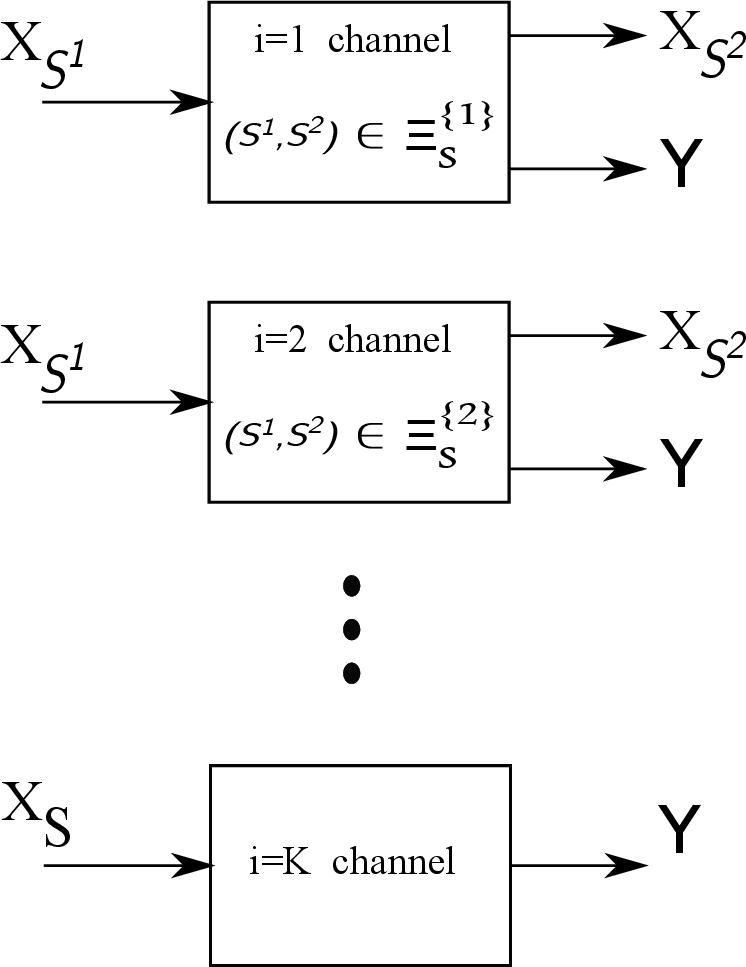}
\caption{Equivalent channel model. Each channel accounts for the case, where $K-i$ of the defective items are recovered and $i$ items are still to be recognized. The total error probability in decoding the true defective set depends on the decoding error probability for each channel. $K-i$ represents the overlap between the true defective set and a false candidate set.} 
\label{fig:channel}
\end{figure}

It is worthwhile mentioning that in the classical channel coding problem \cite{gallager,coverbook}, the error probability analysis for Maximum Likelihood or Typical Set Decoding separates well due to the independence of the channel output and every codeword other than the truly transmitted one. However, in the group testing problem, a main difficulty arises from the fact that an arbitrary set of $K$ items and the true defective set could be overlapping. Hence, independence of the output and the collection of codewords indexed by that set does not hold anymore. That required introducing the previous machinery to analyze the error probability. Two ingredients that were key to our analysis are
\begin{enumerate}
\item Separating the error events $E_i$ of misclassifying the defective set in $i$ items
\item For every $i$, we averaged over realizations of ensemble of codewords of every candidate set while holding fixed the partition common to these sets and the true set of defectives.
\end{enumerate}
\noindent {\bf Further Intuition:}
Intuitively, the numerator in Theorem \ref{thm:main_result} represents the number of bits required to enumerate all possible sets with $i$
misclassified items ($i$ out of $K$), and the denominator represents
the amount of information per test if $K-i$ of the defective items are known.
%
One key idea behind the described approach is illustrated in Fig. \ref{fig:channel}. The group testing problem has been mapped to an equivalent multiple channel model. Each channel accounts for the case where $K-i$ of the defective items are recovered and $i$ items are still to be recognized, where $K-i$ represents the overlap between the true defective set and a false candidate set. The total error probability in decoding the true defective set depends on the decoding error probability for each channel. By studying each error channel separately, we upper bound the overall error probability.

The previous result is a simple mutual information expression that can be used to determine the tradeoffs between $K$, $N$, $T$, and noise for
various models as we show in the following sections. We are interested in determining the required number of tests $T$ to achieve an arbitrarily small error probability for the $2$ aforementioned scaling regimes for different models and different performance criteria. The probability $p$ that an item is pooled
in a given test is a test design parameter, so we choose $p=\frac{1}{K}$.
Table \ref{tab:summary} summarizes the scaling results for the considered models for the average error and worst case error criteria. Note
that the number of tests increases by $1/(1-u)^2$ factor for the
dilution model and only by $1/(1-q)$ for the additive noise model
which matches the aforementioned intuition.

\section{Lower Bound: Fano's inequality}
\label{sec,LB}
In this section we also derive lower bounds on the required number of tests using Fano's inequality \cite{coverbook}. We state the following theorem
\begin{theorem}
\label{thm,LB}
For $N$ items and a set $S_{\omega}$ of $K$ defectives, a lower bound on the total number of tests required to recover the defective set when the components of $\emb{X}$ are i.i.d., is given by
\begin{align}
T\geq\max_{i:({\cal S}^1,{\cal S}^2)\in\Xi_{S_\omega}^{\{i\}}}\frac{\log\binom{N-K+i}{i}}{I(X_{{\cal S}^1};X_{{\cal S}^2},Y)},~~~i=1, 2, \ldots K
\label{eq:lb_equation}
\end{align}
where $Y$ is a binary random variable denoting a test outcome, $X_{S_{\omega}}$ is the per-test input indexed by $S_{\omega}$, and the set $\Xi_{S}^{\{i\}}$ is the set of tuples $({\cal S}^1,{\cal S}^2)$ partitioning the set $S$ into disjoint sets ${\cal S}^1$ and ${\cal S}^2$ with cardinalities $i$ and $K-i$, respectively as defined in (\ref{eq:Xi}).
\end{theorem}

\begin{proof}
The tests' outcomes $Y^T$ are probabilistically related to the index $\omega\in{\cal I}=\{1,2,\ldots,\binom{N}{K}\}$. Suppose $K-i$ items ${\cal S}^2$ are revealed to us. From $Y^T$ we estimate the defective set $\omega$. Let the estimate be $\hat{\omega}=g(Y^T)$. Define the probability of error
\[
P_e = \Pr[\hat{\omega}\ne\omega]
\]
If $E$ is a binary random variable that takes the value $1$ in case of an error i.e., if $\hat{\omega}\ne\omega$, and $0$ otherwise. Then using the chain rule of entropies \cite{coverbook}\cite{gallager} we have
\begin{align}
H(E,\omega|Y^T,\emb{X}_{{\cal S}^2})&=H(\omega|Y^T,\emb{X}_{{\cal S}^2})+H(E|\omega,Y^T,\emb{X}_{{\cal S}^2})\nonumber\\
&=H(E|Y^T,\emb{X}_{{\cal S}^2})+H(\omega|E,Y^T,\emb{X}_{{\cal S}^2})
\label{eq:H(E,W|Y,X)}
\end{align}
The random variable $E$ is fully determined given $Y^T$ and $\omega$. It follows that $H(E|\omega,Y^T,\emb{X}_{{\cal S}^2})=0$. Since $E$ is a binary random variable $H(E|Y^T,\emb{X}_{{\cal S}^2})\leq 1$. Consequently we can bound $H(\omega|E,Y^T,\emb{X}_{{\cal S}^2})$ as follows
\begin{align}
H(\omega|E,Y^T,\emb{X}_{{\cal S}^2})&=P(E=0)H(\omega|E=0,Y^T,\emb{X}_{{\cal S}^2})+P(E=1)H(\omega|E=1,Y^T,\emb{X}_{{\cal S}^2})\nonumber\\
&\leq(1-P_e)0+P_e\log\left(\binom{N-K+i}{i}-1\right)\nonumber\\
&\leq P_e\log\binom{N-K+i}{i}
\end{align}
The second inequality follows from the fact that revealing $K-i$ items, and given that $E=1$, the conditional entropy can be upper bounded by the logarithm of the number of outcomes. From (\ref{eq:H(E,W|Y,X)}), we obtain the genie aided Fano's inequality
\begin{align}
H(\omega|Y^T,\emb{X}_{{\cal S}^2})\leq 1+P_e\log\binom{N-K+i}{i}.
\label{eq:fano}
\end{align}

Since the set ${\cal S}^2$ of $K-i$ defectives is revealed, $\omega$ is uniformly distributed over the set of indices that correspond to subsets of size $K$ containing ${\cal S}^2$. It follows that

\begin{align}
\log\binom{N-K+i}{i}&=H(\omega|\emb{X}_{{\cal S}^2})=H(\omega|Y^T,\emb{X}_{{\cal S}^2})+I(\omega;Y^T|\emb{X}_{{\cal S}^2})\nonumber\\
&\leq 1+P_e\log\binom{N-K+i}{i}+I(\emb{X}_{S_{\omega}};Y^T|\emb{X}_{{\cal S}^2})
\end{align}
Since $S_{\omega}={\cal S}^1\cup{\cal S}^2$, where $({\cal S}^1,{\cal S}^2)\in\Xi_{S_\omega}^{\{i\}}$, we have
\begin{align}
P_e\geq 1-\frac{I(\emb{X}_{{\cal S}^1};Y^T|\emb{X}_{{\cal S}^2})+1}{\log\binom{N-K+i}{i}}
\end{align}

Thus, for the probability of error to be asymptotically bounded away from zero, it is necessary that
\begin{align}
\log\binom{N-K+i}{i}\leq I(\emb{X}_{{\cal S}^1};Y^T|\emb{X}_{{\cal S}^2})
\end{align}

Following a standard set of inequalities we have
\begin{align}
\log\binom{N-K+i}{i}&\leq I(\emb{X}_{{\cal S}^1};Y^T|\emb{X}_{{\cal S}^2})\nonumber\\
&=H(Y^T|\emb{X}_{{\cal S}^2})-H(Y^T|\emb{X}_{S_{\omega}})\nonumber\\
&\buildrel\rm (a)\over=\sum_{t=1}^T H(Y(t)|Y^{t-1},\emb{X}_{{\cal S}^2})- H(Y(t)|X_{S_{\omega}}(t))\nonumber\\
&\buildrel\rm (b)\over\leq \sum_{t=1}^{T} H(Y(t)|X_{{\cal S}^2}(t))-H(Y(t)|X_{S_{\omega}}(t))\nonumber\\
&= \sum_{t=1}^T I(X_{S_{\omega}}(t);Y(t)|X_{{\cal S}^2}(t))\nonumber\\
&= \sum_{t=1}^T I(X_{{\cal S}^1}(t);Y(t)|X_{{\cal S}^2}(t))\nonumber\\
&\buildrel\rm (c)\over= T I(X_{{\cal S}^1};Y|X_{{\cal S}^2})\nonumber\\
&\buildrel\rm (d)\over= T I(X_{{\cal S}^1};X_{{\cal S}^2},Y)
\label{eq:fano_seq_ineqs}
\end{align}
In (a) we made use of the chain rule for entropy and the memoryless property of the channel and (b) is true since conditioning reduces entropy. (c) is due to the i.i.d. assumption. Finally, (d) follows from the chain rule for mutual information and independence, i.e.,
\begin{align}
I(X_{{\cal S}^1};X_{{\cal S}^2},Y)=I(X_{{\cal S}^1};X_{{\cal S}^2})+I(X_{{\cal S}^1};Y|X_{{\cal S}^2})
\end{align}
and $I(X_{{\cal S}^1};X_{{\cal S}^2})=0$.

Since (\ref{eq:fano_seq_ineqs}) has to be true for all $i$, a necessary condition on the total number of tests is given by
\begin{align}
T\geq\max_{i:({\cal S}^1,{\cal S}^2)\in\Xi_{S_\omega}^{\{i\}}}\frac{\log\binom{N-K+i}{i}}{I(X_{{\cal S}^1};X_{{\cal S}^2},Y)}
\end{align}
proving theorem \ref{thm,LB}.
\end{proof}
%
%
\begin{remark}
Note that the bound in Theorem \ref{thm:main_result} is achievable when $N\rightarrow\infty$ for every fixed $K$. As pointed out earlier, for the case where both $N$ and $K$ scale simultaneously with $K=o(N)$, which parallels the lower bound in Theorem \ref{thm,LB}, the achievable bound has an extra polylog factor in $K$. Unlike the lower bound in Theorem \ref{thm,LB}, the sufficiency result is an order scaling. However, the mutual information expression in the denominator in Theorem \ref{thm,LB} matches the mutual information expression in the achievable bound in Theorem \ref{thm:main_result}. Furthermore, since the combinatorial terms in the numerators have a similar asymptotic scaling, the achievable bound and the lower bound are order-wise tight in the asymptotic regime considered in this paper up to a small polylog factor.
%
\end{remark}

\begin{remark}
According to this lower bound, it is not hard to see that for the noise free case $T\geq C\cdot K\log\left(\frac{N}{K}\right)$, for some positive constant $C$, is a necessary condition on the total number of tests.
\end{remark}

%

\section{Noise free case-deterministic output}
\label{sec,noiseless} In this section, we consider the noise-free
(deterministic) case: the test outcome $Y$ is $1$ if and only if a
defective item is pooled in that test. Hence $Y$ is given by (\ref{eq,noise_free}). We consider two scenarios: average and
worst-case error. As mentioned earlier, the former requires the
average error to be small and the latter considers the worst-case
input since bounding the average error probability does not
guarantee error-free performance for all possible defective sets.

\subsection{Average Error Probability}
\begin{theorem}
For $N$ items and $K$ defectives, an arbitrarily small average error probability $P_e$ is achievable for some $T=O(K\log N)$, where $T$ is the total number of tests. In other words, there is a constant $C>0$ independent of $N$ and $K$ such that if $T\geq C\cdot K\log N$ then the probability of error goes to zero.
\label{thm:avgPe_scaling}
\end{theorem}

\begin{proof}
In the noise free case, if $S$ is the defective set, then $H(Y|X_S)=0$, where $H(Y|X_S)$ is the conditional entropy of $Y$ given $X_S$ \cite{coverbook}. The conditional entropy $H(Y|X)$ of a random variable $Y$ given another random variable $X$ is formally defined as the negative of the expected value of the logarithm of the conditional probability $p(Y|X)$, i.e.,
\[
H(Y|X) = -E\log(p(Y|X)
\]
and hence $H(Y|X_S)$ is a measure of uncertainty of the test outcome $Y$ given the input $X_S$ corresponding to the defective set $S$. In other words, since we are dealing with a noise-free channel, $Y$ is determined with certainty if we know that $X_S$ was ``transmitted". For a given $i$ and any $({\cal S}^1,{\cal S}^2)\in\Xi_S^{\{i\}}$, the mutual information expression $I(X_{{\cal S}^1};X_{{\cal S}^2},Y)$ can be written as
\begin{eqnarray}
I(X_{{\cal S}^1};X_{{\cal S}^2},Y)&=&H(Y|X_{{\cal S}^2})-H(Y|X_S)\nonumber\\
&=&(1-p)^{K-i}H((1-p)^i)\nonumber\\
&\geq& \left(1-\frac{1}{K}\right)^{K-i}\left(1-\frac{1}{K}\right)^i \log\frac{1}{(1-\frac{1}{K})^i}
\end{eqnarray}
Thus, for large $K$,
\[
I(X_{{\cal S}^1};X_{{\cal S}^2},Y)\geq e^{-1}\frac{i}{K\ln2}=\Theta\left(\frac{i}{K}\right)
\]
Now, the numerator in Theorem \ref{thm:main_result} is upper bounded by $\Theta(i\log N)$.
Hence, from Theorem \ref{thm:main_result}, $T=O(K\log N)$, i.e., $P_e\rightarrow 0$ for $T\geq C\cdot K\log N$ for some constant $C>0$.
\end{proof}
The following theorem establishes a sufficient condition on the number of tests for the more general case of Section \ref{sec:general_case} when both $N$ and $K$ scale simultaneously with $K=o(N)$.
\begin{theorem}
\label{thm:noisless_appdx}
Consider the noiseless channel in (\ref{eq,noise_free}). There exists a positive constant $C$ independent of $N$ and $K$ such that if
\[
T \geq C\cdot K \log N\log^2 K
\]
then it follows that the probability of misidentifying the K defectives goes to zero when both $N$ and $K$ scale simultaneously with $K=o(N)$.
\end{theorem}
This amounts to an extra $\log^2 K$ factor in the final scaling. This polylog factor introduces some conservatism but appears difficult to avoid based on the bounding techniques that we employ for bounding the error exponents.
The proof of Theorem \ref{thm:noisless_appdx} is presented in Appendix C.

\subsection{Maximum Probability of error}
\label{subsec:max_Pe}
The previous analysis considered the average error case. Maintaining the average error probability below $\epsilon$ is not enough if we are interested in the worst-case set of defectives, i.e., the maximum error case. For exact reconstruction, the worst-case error is required to be zero.

\begin{theorem}
For $N$ items and $K$ defectives, $T=O(K^2\log N)$ is achievable for exact reconstruction (worst-case error criteria).
\end{theorem}

\begin{proof}
The average error probability is below $\epsilon$. Thus,
\begin{equation}
P_e=\sum_{\emb{X}}\Pr[{\emb{X}}]\lambda({\emb{X}})<\epsilon\rightarrow\exists {\emb{X}}:\lambda(\emb{X})<\epsilon.
\end{equation}
In other words, since the average probability of error (over codebooks and inputs) is below $\epsilon$, then there exist a codebook $\emb{X}$ such that $\lambda({\emb{X}})=\frac{1}{\binom{N}{K}}\sum_v \lambda_v({\emb{X}})<\epsilon$. Choosing $\epsilon=\frac{1}{\binom{N}{K}}$ guarantees that the worst case error is also $0$ since for the noiseless case $\lambda_v\in\{0,1\}$.

If $f(\rho)$ denotes the exponent of the error probability $P(E_i)$ in $(\ref{eq:error_exp})$, and noting that $f(0)=0$, we get by the Lagrange form of the Taylor series expansion in the neighborhood of zero that
\[
f(\rho) = \rho\frac{df}{d\rho} + O(\rho^2)
\]
where
\[
\frac{df}{d\rho} = I(X_{{\cal S}^1};X_{{\cal S}^2},Y) - \frac{\log\binom{N-K}{i}\binom{K}{i}}{T}
\]
Hence, the existence of a good codebook combined with the conditions $P(E_i) < \frac{1}{\binom{N}{K}}, \forall i$, ensure perfect reconstruction, i.e.,
\begin{align}
2^{-T(\rho(I(X_{{\cal S}^1};X_{{\cal S}^2},Y) - \frac{\log\binom{N-K}{i}\binom{K}{i}}{T})+O(\rho^2))}<\frac{1}{\binom{N}{K}}
\label{eq:perfect_reconstruction}
\end{align}
which asymptotically translates to the sufficient condition
\[
\Omega\left(\frac{i}{K}\right)-\frac{\Theta(i\log N)}{T}-\frac{\Theta(K\log N)}{T}>0
\]
and some $T=O(K^2\log N)$ is achievable. Note that we have ignored the $O(\rho^2)$ term in deriving this result. Nevertheless, it follows from Lemma \ref{thm:errexp} in Appendix C that this term is not dominant and consequently the result follows. 

\end{proof}

\subsection{Achievability with Distortion}
\label{sec,distortion}
In this section we relax our goal. We are satisfied with recovering a large fraction of the defective items. In other words, we allow an approximate reconstruction \cite{gilbert} in the sense that if $K$ candidates are declared, up to $\alpha K$ misses are allowed ($\alpha$ small).
\begin{theorem}
If $N$ is the total number of items and $K$ the size of the defective set, approximate reconstruction of the defective set, i.e., with up to $\alpha K$ misses, is achievable with some $T=O(K\log N)$.
\end{theorem}

\begin{proof}
In this case, $P_e\leq\sum_{i=\alpha K}^K P(E_i)$. Using an argument analogous to (\ref{eq:perfect_reconstruction}), we verify the error exponent of $P(E_i)$. Since the smallest $i$ is $\Theta(K)$, the sufficient condition on $T$ becomes
\[
\frac{\Theta(K\log N)}{T} > \mbox{constant}
\]
Hence, $T = O(K\log N)$.
\end{proof}


\section{Noisy Group Testing}
\label{sec,noisy} The derived upper bound in Thm.
\ref{thm:main_result} is fairly general as it maps the group testing
problem to an equivalent channel model. This does not restrict the
model to the noise-free scenario and hence could also be used to
account for different noisy versions of the problem. The question
is how easy it is to compute the mutual information
expression. In this section, we derive sufficient conditions on the
number of tests for two types of noisy channels. It is to be noted
that the result could also be applied to other noise models.

\subsection{Additive Observation Noise}
First, we consider the additive output model of Eq.\ref{eq,additive_noise}, i.e.,
\[
Y(t)=\left(\bigvee_{i=1}^N X_i(t)R_i\right)\vee W(t)
\]
where $W$ is Bernoulli($q$). This model captures the
possibility of having false alarms. This accounts for blood tests with false alarms or background wireless losses \cite{Asilomar08} etc.
%
The following theorem provides a sufficient condition on $T$ for the additive noise model.
\begin{theorem}
For the additive noise model in (\ref{eq,additive_noise}), $N$
items, $K$ defectives, there exists $T=O\left(\frac{K\log N}{1-q}\right)$ for which the average error probability $P_e$ goes to zero as $N\rightarrow\infty$ for every fixed $K$, where $q$ is the parameter of the bernoulli distribution of the binary noise.
\label{thm:noisy_additive}
\end{theorem}

{\bf Note:} As $q$ increases, the number of tests required to identify the defective set increases since the outcome of a pooling test becomes less reliable due to false alarms.
\begin{proof}
Consider any partition $({\cal S}^1,{\cal S}^2)\in\Xi_S^{\{i\}}$. Unlike the noise-free model, the conditional entropy $H(Y|X_S)$ is no longer zero. Although, the test outcome is certainly positive if $X_S$ is anything but the all-zero vector, the test outcome is uncertain if $X_S$ is all zero due to the additive noise. Then
\begin{align}
I(X_{{\cal S}^1};X_{{\cal S}^2},Y)&=H(Y|X_{{\cal S}^2})-H(Y|X_S)\nonumber\\
&=(1-p)^{K-i}H\left((1-p)^i(1-q)\right)-(1-p)^{K}H(q)\nonumber\\
&=(1-p)^{K-i}\left[H\left((1-p)^i(1-q)\right)-(1-p)^{i}H(q)\right]
\label{eq:I_additive_step}
\end{align}
The first entropy term in (\ref{eq:I_additive_step}) can be written as
\begin{align}
&H\left((1-p)^i(1-q)\right)=(1-p)^i(1-q)\log\frac{1}{(1-p)^i(1-q)}+(1-(1-p)^i(1-q))\log\frac{1}{1-(1-p)^i(1-q)}\nonumber\\
&=i\left(1-\frac{1}{K}\right)^i(1-q)\log\left(\frac{1}{1-\frac{1}{K}}\right)+\left(1-\frac{1}{K}\right)^i(1-q)\log\left(\frac{1}{1-q}\right) +\sum_{j=1}^{\infty}\frac{1}{j\ln 2}\left(1-\frac{1}{K}\right)^{ji}(1-q)^j\nonumber\\
&\qquad\qquad\qquad-\frac{1}{\ln 2}\left(1-\frac{1}{K}\right)^i(1-q)\left[\sum_{j=1}^{\infty}\frac{1}{j}\left(1-\frac{1}{K}\right)^{ji}(1-q)^j\right]
\label{eq:first_entropy}
\end{align}
where the last equality is obtained through simple Taylor series expansion. Expanding the second entropy term in (\ref{eq:I_additive_step}) we get
\begin{align}
(1-p)^iH(q)=\left(1-\frac{1}{K}\right)^iq\log\frac{1}{q}+\left(1-\frac{1}{K}\right)^i(1-q)\log\frac{1}{1-q}
\label{eq:second_entropy}
\end{align}
Subtracting (\ref{eq:first_entropy}) and (\ref{eq:second_entropy}), and multiplying by $(1-p)^{K-i}$, the mutual information expression simplifies to
\begin{align}
I(X_{{\cal S}^1};X_{{\cal S}^2},Y)&=\left(1-\frac{1}{K}\right)^{K-i}\Bigg[\frac{i}{\ln 2}\left(1-\frac{1}{K}\right)^i(1-q)\left(\sum_{j=1}^{\infty}\frac{1}{jK^j}\right)+\frac{1}{\ln 2}\left(1-\frac{1}{K}\right)^i(1-q)\nonumber\\
&-\frac{1}{\ln 2}\left(\sum_{j=2}^{\infty}\frac{1}{j(j-1)}\left(1-\frac{1}{K}\right)^{ji}(1-q)^j\right)
-\left(1-\frac{1}{K}\right)^iq\log\frac{1}{q}\Bigg]\nonumber\\
&\geq \frac{i}{\ln 2}\left(1-\frac{1}{K}\right)^K(1-q)\left(\sum_{j=1}^{\infty}\frac{1}{jK^j}\right)+\frac{1}{\ln 2}\left(1-\frac{1}{K}\right)^K(1-q)\nonumber\\
&-\frac{1}{\ln 2}\left(1-\frac{1}{K}\right)^K\left[\sum_{j=2}^{\infty}\frac{(1-q)^j}{j(j-1)}\right]-\left(1-\frac{1}{K}\right)^Kq\log\frac{1}{q}
\label{eq:bound_I}
\end{align}
Now note that
\begin{align}
\sum_{j=2}^{\infty}\frac{(1-q)^j}{j(j-1)}=1-q-q\ln2\log\frac{1}{q}
\label{eq:log_simplification}
\end{align}
This is not hard to verify since
\begin{eqnarray}
\frac{(1-q)^2}{2}+\frac{(1-q)^3}{6}+\ldots&=&(1-q)^2\left(1-\frac{1}{2}\right)+(1-q)^3\left(\frac{1}{2}-\frac{1}{3}\right)+\ldots\nonumber\\
&=&(1-q)\left(1-q+\frac{(1-q)^2}{2}+\ldots\right)-\left(\frac{(1-q)^2}{2}+\frac{(1-q)^3}{3}+\ldots\right)\nonumber\\
&=&(1-q)\ln2\log\frac{1}{q}-\left(\ln2\log\frac{1}{q}-(1-q)\right)\nonumber\\
&=& 1-q-q\ln2\log\frac{1}{q}
\end{eqnarray}
Replacing in (\ref{eq:bound_I}) we get
\begin{align}
I(X_{{\cal S}^1};X_{{\cal S}^2},Y)&\geq\frac{i}{K\ln 2}\left(1-\frac{1}{K}\right)^K(1-q)
\label{eq:I_additive}
\end{align}
Following the same argument in the proof of theorem \ref{thm:avgPe_scaling}, replacing (\ref{eq:I_additive}) in Theorem \ref{thm:main_result}, we have that $T\leq\frac{K\ln 2\log N}{1-q}$, i.e., $T=O\left(\frac{K\log N}{1-q}\right)$ is achievable.
\end{proof}
The same scaling holds for the regime where $N$ and $K$ simultaneously approach infinity with $K=o(N)$. We state the result in the following theorem and present the proof in Appendix C.
\begin{theorem}
For the additive noise model in (\ref{eq,additive_noise}), $N$
items, $K$ defectives, there exists $T=O\left(\frac{K\log N}{1-q}\right)$ for which the average error probability of misidentifying the $K$ items goes to zero as $N,K\rightarrow\infty$ with $K=o(N)$, where $q$ is the parameter of the bernoulli distribution of the binary noise.
\label{thm:additive_general}
\end{theorem}
This follows immediately from Lemma \ref{lem:additive_Eo2} and the Lagrange form of the Taylor series expansion of the error exponent in Lemma \ref{thm:errexp} in Appendix C.

\begin{remark}
Following the same argument in Section \ref{subsec:max_Pe}, it is
not hard to see that the same scaling holds for the worst case error
criteria but replacing $K$ with $K^2$ as shown in table
\ref{tab:summary}. Recall that the worst-case error criterion implied exact recovery in the noise-free model. Exact recovery emerged from the fact that if the average probability of error using a randomly generated codebook is bounded from above by some $\epsilon$, then there exist a \emph{good} codebook as argued in Section \ref{subsec:max_Pe}. Hence choosing such a ``good" codebook by design, and by proper choice of $\epsilon$, we could guarantee exact recovery for the worst case input in the noise-free model. However, in this noisy setup, the worst case error criteria only means that the
worst-case error probability goes to zero in contrast to the
noise-free case where this scaling ensured exact reconstruction of the defective set.
\end{remark}

The next theorem establishes a necessary condition on the number of tests for the additive noise model.

\begin{theorem}
Considering the additive noise model in (\ref{eq,additive_noise}), a total number of items $N$ and $K$ defectives, a necessary condition for the number of tests is that
\[
T=\Omega\left(\frac{K\log \frac{N}{K}}{2(1-q)+\ln(\frac{1}{q})}\right)
\]
where $q$ is the parameter of the bernoulli distribution of the binary noise. Namely, if  $T =o\left(\frac{K\log \frac{N}{K}}{2(1-q)+\ln(\frac{1}{q})}\right)$ then the error probability approaches 1.
\label{thm:noisy_additive_LB}
\end{theorem}

\begin{proof}
See Appendix B
\end{proof}

\subsection{Dilution}
The second noisy model we consider is the ``dilution" model.
Positive items taking part in a given test might probabilistically
behave as absent (diluted). If all positive items in a given test
appear as absent that could potentially lead to erroneously zero
outcomes. This model is motivated by blood dilution due to pooling
with other negative tests, or imperfectly diluted blood samples, or
adversarial camouflage-in the form of probabilistic transmission-in
communication systems \cite{Asilomar08}. This is captured by the
Z-channel model of Fig. \ref{fig,z_channel}. For this case, we show
that there exists $T=O(\frac{K\log N}{(1-u)^2})$ which is achievable. Intuitively, a larger flip probability $u$ implies that more items will get
diluted. As the tests become less reliable, a larger number of tests
is needed to identify the defective set.


\begin{theorem}
Considering $N$ items, $K$ defectives and the dilution model
represented by the Z-channel in Fig.(\ref{fig,z_channel}) and
(\ref{eq:dilution_model}), for some number of tests $T=O\left(\frac{K\log N}{(1-u)^2}\right)$ the average error probability asymptotically approaches zero, where $u$ is the transition probability of the Z-channel (i.e. the probability that $1$ is flipped into $0$).
\label{thm,dilution}
\end{theorem}

\begin{proof}
For notational convenience we let $u=1-s$. Again, consider a partition $({\cal S}^1,{\cal S}^2)\in\Xi_S^{\{i\}}$
\begin{align}
I(X_{{\cal S}^1};X_{{\cal S}^2},Y)&=H(Y|X_{{\cal S}^2})-H(Y|X_S)\nonumber\\
&=\sum_{j=0}^{K-i}\binom{K-i}{j}\left(\frac{1}{K}\right)^j\left(1-\frac{1}{K}\right)^{K-i-j}H\left(\sum_{\ell=0}^i\binom{i}{\ell}(1-s)^{j+\ell}\left(\frac{1}{K}\right)^{\ell}\left(1-\frac{1}{K}\right)^{i-\ell}\right)\nonumber\\
&- \sum_{j=1}^{K}\binom{K}{j}\left(\frac{1}{K}\right)^j\left(1-\frac{1}{K}\right)^{K-j}H\left((1-s)^j\right)
\label{eq:I_dilution_step}
\end{align}
The first sum i.e., $H(Y|X_{{\cal S}^2})$, can be written as
\begin{align}
H(Y|X_{{\cal S}^2})&=\sum_{j=0}^{K-i}\binom{K-i}{j}\left(\frac{1}{K}\right)^j\left(1-\frac{1}{K}\right)^{K-i-j}H\left((1-s)^j(1-\frac{s}{K})^i\right)\nonumber\\
&=\sum_{j=0}^{K-i}\binom{K-i}{j}\left(\frac{1}{K}\right)^j\left(1-\frac{1}{K}\right)^{K-i-j}\left[(1-s)^j(1-\frac{s}{K})^i\log\frac{1}{(1-s)^j(1-\frac{s}{K})^i}\right]\nonumber\\
&+\sum_{j=0}^{K-i}\binom{K-i}{j}\left(\frac{1}{K}\right)^j\left(1-\frac{1}{K}\right)^{K-i-j}\left[\left(1-(1-s)^j(1-\frac{s}{K})^i\right)\log\frac{1}{1-(1-s)^j(1-\frac{s}{K})^i}\right]\nonumber\\
&=\eta_1+\eta_2
\label{eq:term1_dilution}
\end{align}
Notice that
\begin{equation}
\sum_{j=0}^{K-i}\binom{K-i}{j}\left(\frac{1}{K}\right)^j\left(1-\frac{1}{K}\right)^{K-i-j}(1-s)^j=\left(1-\frac{s}{K}\right)^{K-i}
\label{eq:z_transform}
\end{equation}

\begin{equation}
\sum_{j=0}^{K-i} j\cdot\binom{K-i}{j}\left(\frac{1}{K}\right)^j\left(1-\frac{1}{K}\right)^{K-i-j}(1-s)^j=\frac{1}{K}(K-i)(1-s)\left(1-\frac{s}{K}\right)^{K-i-1}
\label{eq:diff_z_transform}
\end{equation}
Using (\ref{eq:z_transform}) and (\ref{eq:diff_z_transform}), the first term $\eta_1$ in (\ref{eq:term1_dilution}) simplifies to
\begin{equation}
\eta_1=i(1-\frac{s}{K})^K\log(\frac{1}{1-\frac{s}{K}})+\frac{1}{K}(K-i)(1-\frac{s}{K})^{K-1}(1-s)\log\frac{1}{1-s}
\end{equation}

The second sum in (\ref{eq:I_dilution_step}) i.e., $H(Y|X_S)$, simplifies to
\begin{align}
H(Y|X_S)&=\sum_{j=0}^{K}\binom{K}{j}\left(\frac{1}{K}\right)^j\left(1-\frac{1}{K}\right)^{K-j}(1-s)^j\log\frac{1}{(1-s)^j}\nonumber\\
&+\sum_{j=0}^{K}\binom{K}{j}\left(\frac{1}{K}\right)^j\left(1-\frac{1}{K}\right)^{K-j}(1-(1-s)^j)\log\frac{1}{1-(1-s)^j}\nonumber\\
&=\theta_1+\theta_2
\label{eq:term2_dilution}
\end{align}
From (\ref{eq:diff_z_transform}), the term $\theta_1$ simplifies to
\begin{equation}
\theta_1=\frac{1}{K}K(1-s)\left(1-\frac{s}{K}\right)^{K-1}\log\frac{1}{1-s}
\end{equation}
Combining $\eta_1$ and $\theta_1$ we get
\begin{eqnarray}
\eta_1-\theta_1&=&i\left(1-\frac{s}{K}\right)^{K-1}\left[\left(1-\frac{s}{K}\right)\log\frac{1}{1-\frac{s}{K}}-\frac{1}{K}(1-s)\log\frac{1}{1-s}\right]\nonumber\\
&=&\frac{i}{\ln2}\left(1-\frac{s}{K}\right)^{K-1}\left[\frac{s}{K}+O\left(\frac{1}{K^2}\right)-\frac{1}{K}(1-s)\left(s+\frac{s^2}{2}+\ldots\right)\right]\nonumber\\
&=&\frac{i}{\ln2}\left(1-\frac{s}{K}\right)^{K-1}\left[\frac{s^2}{K}\left(1-\frac{1}{2}\right)+\frac{s^3}{K}\left(\frac{1}{2}-\frac{1}{3}\right)+\ldots\right]\nonumber\\
&\geq& \frac{is^2}{2K\ln2}\left(1-\frac{s}{K}\right)^{K-1}
\label{eq:eta1_minus_theta1}
\end{eqnarray}
Now we consider the remaining terms i.e., $\eta_2$ and $\theta_2$. It is sufficient to show that the difference $\eta_2-\theta_2$ is $\geq 0$. First consider the term $\eta_2$ in (\ref{eq:term1_dilution})
\begin{align}
\eta_2=\sum_{j=0}^{K-i}\binom{K-i}{j}\left(\frac{1}{K}\right)^j\left(1-\frac{1}{K}\right)^{K-i-j}\left[\left(1-(1-s)^j\left(1-\frac{s}{K}\right)^i\right)\log\frac{1}{1-(1-s)^j(1-\frac{s}{K})^i}\right]\nonumber
\end{align}
\begin{align}
=\frac{1}{\ln2}\sum_{j=0}^{K-i}\binom{K-i}{j}\left(\frac{1}{K}\right)^j\left(1-\frac{1}{K}\right)^{K-i-j}&\left[1-(1-s)^j\left(1-\frac{s}{K}\right)^i\right] \Big[\sum_{\ell=1}^{\infty}\frac{1}{\ell}(1-s)^{\ell j}\left(1-\frac{s}{K}\right)^{\ell i}\Big]
\end{align}
From (\ref{eq:z_transform}), $\eta_2$ simplifies to:
\begin{eqnarray}
\eta_2&=&\frac{1}{\ln 2}\left[\left(1-\frac{s}{K}\right)^K-\sum_{j=0}^{K-i}\binom{K-i}{j}\left(\frac{1}{K}\right)^j\left(1-\frac{1}{K}\right)^{K-i-j}\left(\sum_{\ell=2}^{\infty}\frac{1}{\ell(\ell-1)}(1-s)^{\ell j}\left(1-\frac{s}{K}\right)^{\ell i}\right)\right]\nonumber\\
&=&\frac{1}{\ln 2}\left[\left(1-\frac{s}{K}\right)^K-\sum_{\ell=2}^{\infty}\frac{1}{\ell(\ell-1)}\left(1-\frac{s}{K}\right)^{\ell i}\left(1-\frac{1}{K}+\frac{1}{K}(1-s)^{\ell}\right)^{K-i}\right]
\label{eq:eta2}
\end{eqnarray}
Now consider the term $\theta_2$ in (\ref{eq:term2_dilution}),
\begin{eqnarray}
\theta_2&=&\frac{1}{\ln 2}\left[\left(1-\frac{s}{K}\right)^K-\sum_{j=0}^{K}\binom{K}{j}\left(\frac{1}{K}\right)^j\left(1-\frac{1}{K}\right)^{K-j}\left(\sum_{\ell=2}^{\infty}\frac{1}{\ell(\ell-1)}(1-s)^{\ell j}\right)\right]\nonumber\\
&=&\frac{1}{\ln 2}\left[\left(1-\frac{s}{K}\right)^K-\sum_{\ell=2}^{\infty}\frac{1}{\ell(\ell-1)}\left(1-\frac{1}{K}+\frac{1}{K}(1-s)^{\ell}\right)^{K}\right]
\label{eq:theta2}
\end{eqnarray}
Comparing (\ref{eq:eta2}) and (\ref{eq:theta2}) it is now clear that for large K, the difference $\eta_2-\theta_2\geq 0$. This is easy to verify since
\begin{align*}
\left(1-\frac{1}{K}+\frac{1}{K}(1-s)^{\ell}\right)^K=
\left(1-\frac{1}{K}+\frac{1}{K}(1-s)^{\ell}\right)^{K-i} \left(1-\frac{1}{K}+\frac{1}{K}(1-s)^\ell\right)^i
\end{align*}
and we only need to verify that
\begin{align}
\left(1-\frac{s}{K}\right)^{\ell}<1-\frac{1}{K}+(1-s)^{\ell}
\label{eq:eta_2_theta_2}
\end{align}
which is obviously true for large $K$, since the LHS goes to $1$ and the RHS goes to $1+(1-s)^{\ell}$ in the limit. Thus, some $T=O(\frac{K\log N}{s^2})$ is achievable. Replacing for $s=1-u$, Theorem \ref{thm,dilution} follows.
\end{proof}
The following theorem provides a sufficient condition on the number of tests for the more general case of Section \ref{sec:general_case} when both $N$ and $K$ scale simultaneously with $K=o(N)$.
\begin{theorem}
\label{thm:dilution_appdx}
Consider the dilution channel model in (\ref{eq:dilution_model}). There exists a positive constant $C$ independent of $N$ and $K$ such that if
\[
T \geq C\cdot K \log N\log^2 K
\]
then it follows that the probability of misidentifying the K defectives goes to zero when both $N$ and $K$ scale simultaneously with $K=o(N)$.
\end{theorem}
Hence, the same result with an extra $\log^2 K$ factor holds for the asymptotic regime where $N$ and $K$ simultaneously approach infinity with $K=o(N)$. This follows immediately from Lemma \ref{lem:dilution_Eo2} and the Lagrange form of the Taylor series expansion of the error exponent in Lemma \ref{thm:errexp} in Appendix C. Again, we point out that this polylog factor introduces some
conservatism but appears difficult to avoid based on the bounding techniques that we employ for bounding the error exponents.

\begin{remark}
Following the same argument in Section \ref{subsec:max_Pe} using (\ref{eq:eta1_minus_theta1}) and Lemma \ref{lem:dilution_Eo2} it is not hard to see that the same scaling holds for the worst-case error criteria but replacing $K$ with $K^2$ as shown in table \ref{tab:summary}.
\end{remark}

\section{Conclusions}
\label{sec,conclusions}
In this paper, we adopted a new information theoretic framework to address group testing problems. This approach shifts the philosophy of random disjunct/separable matrix generation to an equivalent channel model and capacity computation. The result is a fairly general achievable bound that enables us to obtain the required tradeoffs between the number of tests, number of items and number of defective items for a wide range of group testing problems. Obtaining these tradeoffs reduces to a simple computation of mutual information expressions. We obtain the asymptotic scaling for i) noise-free setups with average and worst-case errors; ii) Approximate reconstruction and iii) Noisy versions of group testing, namely additive and dilution models. We establish that the sufficiency result is tight as it matches (order-wise) a necessary condition on the total number of tests.

\renewcommand{\theequation}{A.\arabic{equation}}
\begin{center} {\bf Appendix A: Proof of Lemmas and Theorems of Section \ref{sec,main_result}}
\end{center}  \setcounter{equation}{0}

\subsection{Proof of Equation~(\ref{eq.ratio})}
Let $\zeta_{\omega}$, $\omega\in{\cal A}$ denote the event where $\omega$ is more likely than $1$. Then, from the definition of ${\cal A}$, the $2$ encoded messages differ in $i$ codewords. Hence
\begin{align}
\Pr[E_i | \omega_0 = 1, \emb{X}_{S_1}, Y^T]&\leq P(\bigcup_{\substack{\omega\in{\cal A}}} \zeta_{\omega}) \leq \sum_{\substack{\omega\in{\cal A}}}P(\zeta_{\omega}) \nonumber
\end{align}
Now note that $\emb{X}_{S_1}$ shares $(K-i)$ codewords with $\emb{X}_{S_{\omega}}$. Following the introduced notation, the common codewords are denoted $\emb{X}_{S_{1,\omega}}$, which is a $(K-i)\times T$ submatrix. The remaining $i$ codewords which are in $\emb{X}_{S_1}$ but not in $\emb{X}_{S_{\omega}}$ are $\emb{X}_{S_{1,\omega^c}}$. Similarly, $\emb{X}_{S_{1^c,\omega}}$ denotes the codewords in $\emb{X}_{S_{\omega}}$ but not in $\emb{X}_{S_1}$. In other words $\emb{X}_{S_1} = (\emb{X}_{S_{1,\omega}};\emb{X}_{S_{1,\omega^c}})$ and $\emb{X}_{S_{\omega}} = (\emb{X}_{S_{1,\omega}};\emb{X}_{S_{1^c,\omega}})$, where the notation $(\emb{F}^{n_1\times T};\emb{G}^{n_2\times T})$ denotes an $(n_1+n_2)\times T$ matrix with a submatrix $\emb{F}$ in the first $n_1$ rows and $\emb{G}$ in the remaining $n_2$ rows. Thus,
\begin{align}
P(\zeta_{\omega}) & = \sum_{\substack{\emb{X}_{S_{\omega}}:p(Y^T|\emb{X}_{S_{\omega}})\geq p(Y^T|\emb{X}_{S_1})}} Q(\emb{X}_{S_{\omega}}|\emb{X}_{S_1})\nonumber\\
& \leq \sum_{\substack{\emb{X}_{S_{1^c,\omega}}}} Q(\emb{X}_{S_{1^c,\omega}})\frac{p(Y^T|\emb{X}_{S_{\omega}})^s}{p(Y^T|\emb{X}_{S_1})^s}~~~~ \forall s>0,~\forall\omega\in{\cal A}
\label{eq:zeta_m}
\end{align}

By independence $Q(\emb{X}_{S_1}) = Q(\emb{X}_{S_{1,\omega}})Q(\emb{X}_{S_{1,\omega^c}})$. Similarly, $Q(\emb{X}_{S_{\omega}}) = Q(\emb{X}_{S_{1,\omega}})Q(\emb{X}_{S_{1^c,\omega}})$. Since we are conditioning on a particular $\emb{X}_{S_1}$, the partition $\emb{X}_{S_{1,\omega}}$ is fixed in the summation in (\ref{eq:zeta_m}) and
\begin{align} \label{eq.ratio1}
P(\zeta_{\omega}) &\leq \sum_{\emb{X}_{S_{1^c,\omega}}} Q(\emb{X}_{S_{1^c,\omega}})\frac{p(Y^T,\emb{X}_{S_{1,\omega}}|\emb{X}_{S_{1^c,\omega}})^s}{Q(\emb{X}_{S_{1,\omega}}|\emb{X}_{S_{1^c,\omega}})^s}\frac{Q(\emb{X}_{S_{1,\omega}}|\emb{X}_{S_{1,\omega^c}})^s}{p(Y^T,\emb{X}_{S_{1,\omega}}|\emb{X}_{S_{1,\omega^c}})^s}\nonumber\\
&\leq \sum_{\emb{X}_{S_{1^c,\omega}}} Q(\emb{X}_{S_{1^c,\omega}})\frac{p(Y^T,\emb{X}_{S_{1,\omega}}|\emb{X}_{S_{1^c,\omega}})^s}{p(Y^T,\emb{X}_{S_{1,\omega}}|\emb{X}_{S_{1,\omega^c}})^s}~~~~ \forall s>0
\end{align}
where the second inequality follows from the independence of the codewords, i.e. $Q(\emb{X}_{S_{1,\omega}}|\emb{X}_{S_{1,\omega^c}}) = Q(\emb{X}_{S_{1,\omega}}|\emb{X}_{S_{1^c,\omega}}) = Q(\emb{X}_{S_{1,\omega}})$.

\subsection{Proof of Lemma~\ref{thm:P(Ei)}}
The main idea of the proof is to:
\begin{itemize}
\item[(1)] Exploit symmetry of the codebook construction.
 \item[(2)] Exploit channel symmetry: The probability kernels $p_{\omega}(Y^T\mid \emb{X}_{S_{\omega}})$ and $p_1(Y^T \mid \emb{X}_{S_1})$, are invariant to permutations of the set $S_{\omega}$ and $S_1$. Specifically, the set $E_{\omega}$ in (\ref{eq:E_omega}) and the associated probability do not depend on $\omega \in {\cal A}$ if for a fixed output, $Y^T$, the codeword values realized on the $S_{1,\omega},\,S_{1,\omega^c},\,S_{1^c,\omega}$ are identical.
\end{itemize}
With this in mind and to avoid confusion we introduce new notation. Let $\emb{W_1}$ and $\emb{W_2}$, denote the values realized on the common part, $S_{1,\omega}$ and the missed part, $S_{1,\omega^c}$ respectively.
Formally, for any $\omega\in{\cal A}$, we let $\emb{W_1}\in{\cal W}_1$ denote a realization of the random submatrix $\emb{X}_{S_{1,\omega}}$ with ${\cal W}_1$ denoting the alphabet. For the binary case, ${\cal W}_1$ is simply $\{0,1\}^{(K-i)\times T}$. Similarly, $\emb{W_2}\in{\cal W}_2$ defines a realization of submatrix $\emb{X}_{S_{1,\omega^c}}$ with ${\cal W}_2=\{0,1\}^{i\times T}$ for the binary case for all $\omega\in{\cal A}$.

Consequently, we note that the joint probability with this new notation $\emb{W_1}$ and $\emb{W_2}$ can be written in terms of the original random vectors, i.e., as the probability that $\emb{X}_{S_{1,\omega}}=\emb{W_1}$ and $\emb{X}_{S_{1,\omega^c}}=\emb{W_2}$ for some fixed $\omega$. So,
\begin{align} \label{eq:realiz}
Prob\{\emb{W_1},\, \emb{W_2}\} &\doteq p(\emb{W_1},\, \emb{W_2}) = Q(\emb{X}_{S_{1,\omega}}=\emb{W_1},\,\emb{X}_{S_{1,\omega}}=\emb{W_2}) \\
Prob\{Y^T \mid \emb{W_1},\, \emb{W_2}\} &\doteq p(Y^T \mid \emb{W_1},\, \emb{W_2}) = p_1(Y^T \mid \emb{W_1},\,\emb{W_2})
\end{align}
An alternative interpretation is to independently pick $K-i$ and $i$ binary vectors of length $T$ for each of the $K-i$ and $i$ common and missed slots respectively. The probability of picking a random vector for any slot is governed by the probability of picking a codeword for that slot.

With this as the notation we now describe the error set. Consider for $\omega \in {\cal A}$ and output $Y^T$ the set of all $\emb{X}$ that result in an error. ${\cal A}$ is defined in ($\ref{eq:setA}$) as the collection of subsets of $\Omega$ which overlap at $K-i$ places with $\omega=1$ and has cardinality $$|{\cal A}|=\log\binom{N-K}{i}\binom{K}{i}.$$
The set of all $\emb{X}$ for a fixed $\omega\in {\cal A}$ and outcomes $Y^T$ is given by:
\begin{align}
E_{\omega} &= \{\emb{X} \mid P(Y^T \mid \emb{X}_{S_{\omega}}) \geq P(Y^T \mid \emb{X}_{S_1})\} \\
&=\{\emb{X} \mid P(Y^T \mid \emb{X}_{S_{1,\omega}},\emb{X}_{S_{1^c,\omega}}) \geq P(Y^T \mid \emb{X}_{S_{1,\omega}},\emb{X}_{S_{1,\omega^c}})\}
\label{eq:E_omega}
\end{align}
Taking the union over all $\omega \in {\cal A}$ we get:
\begin{align}
E = \cup_{\omega \in {\cal A}} E_{\omega}
\end{align}

We next define the set of all $\emb{X}$ such that the codewords take the same value on the common and missed parts, i.e.,
\[
B_{\omega}(\emb{W_1},\emb{W_2})=\{\emb{X}: \emb{X}_{S_{1,\omega}} = \emb{W_1},\,\,\emb{X}_{S_{1,\omega^c}} = \emb{W_2} \},\,\,\, \omega \in {\cal A}
\]
where $\emb{X}_{S_{1,\omega}} = \emb{W_1}$ and $\emb{X}_{S_{1,\omega^c}} = \emb{W_2}$ is component-wise, i.e., the first component of $\emb{X}_{S_{1,\omega}}$ is equal to the first component of $\emb{W_1}$ and so on, where each component refers to a row vector of length $T$.

We now let
\[
E(\emb{W_1},\emb{W_2},Y^T) = \cup_{\omega \in {\cal A}} \left (E_{\omega} \cap B_{\omega}(\emb{W_1},\emb{W_2}) \right )
\]
We next compute the error probability conditioned on the realization of common and missed values, $\emb{W_1},\emb{W_2}$, and the output $Y^T$:
\begin{align*}
Prob\{ E(\emb{W_1},\emb{W_2},Y^T) \mid \emb{W_1},\emb{W_2},Y^T\} &= \sum_{\emb{X} \in E(\emb{W_1},\emb{W_2})} Q(\emb{X}\mid \emb{W_1},\emb{W_2},Y^T) \\
&\leq \sum_{\omega \in {\cal A}}\sum_{{\tiny \begin{array}{c} \emb{X} \in E_{\omega} \\ \emb{X}_{S_{1,\omega}}=\emb{W_1} \\ \emb{X}_{S_{1,\omega^c}}=\emb{W_2} \end{array}}} Q(\emb{X} \mid \emb{W_1},\emb{W_2},Y^T) \\
&= \sum_{\omega \in {\cal A}}\sum_{{\tiny \begin{array}{c} \emb{X} \in E_{\omega} \\ \emb{X}_{S_{1,\omega}}=\emb{W_1} \\ \emb{X}_{S_{1,\omega^c}}=\emb{W_2} \end{array}}} Q(\emb{X} \mid \emb{W_1},\emb{W_2}) \\
&= \sum_{\omega \in {\cal A}}\sum_{{\tiny \begin{array}{c} \emb{X} \in E_{\omega} \\ \emb{X}_{S_{1,\omega}}=\emb{W_1} \\ \emb{X}_{S_{1,\omega^c}}=\emb{W_2} \end{array}}} Q(\emb{X}_{S_{1^c,\omega}},\emb{X}_{S_{1,\omega^c}},\emb{X}_{S_{1^c,\omega}} \mid \emb{W_1},\emb{W_2})\\& =
\sum_{\omega \in {\cal A}}\sum_{{\tiny \begin{array}{c} \emb{X} \in E_{\omega} \\ \emb{X}_{S_{1,\omega}}=\emb{W_1} \\ \nonumber \emb{X}_{S_{1,\omega^c}}=\emb{W_2} \end{array}}} Q(\emb{X}_{S_{1^c,\omega}})
\end{align*}
where the second inequality is not an equality because of possible repetition of codewords. If the codewords for two different common sets $S_{1,\omega_1}$ and $S_{1,\omega_2}$ where $\omega_1,\omega_2 \in {\cal A}$ are identical then we possibly double count by taking the summation over $\omega\in {\cal A}$. The third equality follows from Bayes rule where we have used the fact that $Y^T$ is conditionally independent of codewords when conditioned on the common and missed values $\emb{W_1}$ and $\emb{W_2}$. The fourth step follows from marginalization of rows the codeword matrix $\emb{X}$. The final step follows from independence of the codewords. We now simplify the result through the following steps:
\begin{align}
Prob\{ E(\emb{W_1},\emb{W_2},Y^T) \mid \emb{W_1},\emb{W_2},Y^T\}
&\stackrel{(a)}{\leq} |{\cal A}| \sum_{{\tiny \begin{array}{c} \emb{X} \in E_{\omega} \\ \emb{X}_{S_{1,\omega}}= \emb{W_1} \\ \nonumber \emb{X}_{S_{1,\omega^c}}=\emb{W_2} \end{array}}} Q(\emb{X}_{S_{1^c,\omega}})\\
\nonumber &\stackrel{(b)}{\leq} |{\cal A}| \sum_{\emb{X}_{S_{1^c,\omega}}} Q(\emb{X}_{S_{1^c,\omega}}) {p^s_{\omega}(Y^T\mid \emb{X}_{S_{1,\omega}}=\emb{W_1}, \emb{X}_{S_{1^c,\omega}}) \over p^s_1(Y^T\mid \emb{X}_{S_{1,\omega}}=\emb{W_1}, \emb{X}_{S_{1,\omega^c}}=\emb{W_2})}\\ \nonumber &\stackrel{(c)}{=} |{\cal A}| \sum_{\emb{X}_{S_{1^c,\omega}}} Q(\emb{X}_{S_{1^c,\omega}}) {p^s_{\omega}(Y^T, \emb{W_1} \mid \emb{X}_{S_{1^c,\omega}}) \over p^s_1(Y^T, \emb{W_1} \mid \emb{W_2})}\\
&= |{\cal A}| \sum_{\emb{X}_{S_{1^c,\omega}}} Q(\emb{X}_{S_{1^c,\omega}}) {p^s_1(Y^T, \emb{W_1} \mid \emb{X}_{S_{1^c,\omega}}) \over p^s_1(Y^T, \emb{W_1} \mid \emb{W_2})} \label{eq.condrealiz}
\end{align}
where (a) follows from the fact that the inner sum does not depend on $\omega$; (b) follows as in Equation~\ref{eq:zeta_m};(c) follows as in Eq.~\ref{eq.ratio1} in the proof of Equation~\ref{eq.ratio}. The last step is based on the observation that the channel probabilities are symmetric and do not depend on $\omega$. Next we compute the total error probability. The proof now follows the same approach we used in Section \ref{subsec.mainresult}. It follows that,
\begin{align*}
P(E_i)=Prob\{E\} = \sum_{\emb{W_1},\emb{W_2},Y^T} p_1(\emb{W_1},\emb{W_2},Y^T) Prob(E(\emb{W_1},\emb{W_2})\mid \emb{W_1},\emb{W_2},Y^T)
\end{align*}
Substituting Eq.~\ref{eq.condrealiz} we obtain:
\begin{align*}
P(E_i) \leq \sum_{\emb{W_1},\emb{W_2},Y^T} p_1(\emb{W_1},\emb{W_2},Y^T)\left ( |{\cal A}| \sum_{\emb{X}_{S_{1^c,\omega}}}
Q(\emb{X}_{S_{1^c,\omega}}) {p^s_1(Y^T,\emb{W_1} \mid \emb{X}_{S_{1^c,\omega}})
\over p^s_1(\emb{Y},\emb{W_1} \mid \emb{W_2})}\right )^{\rho}
\end{align*}
Next we substitute Eq.~\ref{eq:realiz} and get the following chain of inequalities:
\begin{align*}
P(E_i) &\leq
   \sum_{\emb{W_1},\emb{W_2},Y^T} Q(\emb{W_2})p_1(Y^T,\emb{W_1} \mid \emb{W_2})\left ( |{\cal A}| \sum_{\emb{X}_{S_{1^c,\omega}}}
Q(\emb{X}_{S_{1^c,\omega}}) {p^s_1(Y^T,\emb{W_1} \mid \emb{X}_{S_{1^c,\omega}})
\over p^s_1(Y^T,\emb{W_1} \mid \emb{W_2})}\right )^{\rho}\\
&=\sum_{\emb{W_1},\emb{W_2},Y^T} Q(\emb{W_2})p^{1-s\rho}_1(Y^T, \emb{W_1} \mid \emb{W_2})\left ( |{\cal A}| \sum_{\emb{X}_{S_{1^c,\omega}}}
Q(\emb{X}_{S_{1^c,\omega}}) p^s_1(Y^T,\emb{W_1} \mid \emb{X}_{S_{1^c,\omega}}) \right )^{\rho}\\
&=\sum_{\emb{W_1},\emb{W_2},Y^T} Q(\emb{W_2})p^{1-s\rho}_1(Y^T,\emb{W_1} \mid \emb{W_2})\left ( |{\cal A}| \sum_{\emb{W_2}} Q(\emb{W_2}) p^s_1(Y^T,\emb{W_1} \mid \emb{W_2}) \right )^{\rho} \\
&=\sum_{\emb{W_1},Y^T} \sum_{\emb{W_2}} Q(\emb{W_2})p^{1-s\rho}_1(Y^T, \emb{W_1} \mid \emb{W_2})\left ( |{\cal A}| \sum_{\emb{W_2}} Q(\emb{W_2}) p^s_1(Y^T,\emb{W_1} \mid \emb{W_2}) \right )^{\rho}\\
&=|{\cal A}|^{\rho}\sum_{\emb{W_1},Y^T} \left ( \sum_{\emb{W_2}} Q(\emb{W_2}) p^{1\over 1+\rho}_1(Y^T,\emb{W_1} \mid \emb{W_2}) \right )^{1+\rho}
\end{align*}
where in the last step we have substituted $s=1/(1+\rho)$. We now observe that $\emb{W_1}$ and $\emb{W_2}$ are dummy variables and so can be replaced with the original symbols $\emb{X}_{S_{1,\omega}}$ and $\emb{X}_{S_{1,\omega^c}}$. Lemma~\ref{thm:P(Ei)} now follows from the last equality by the independence of tests and by direct comparison.

\renewcommand{\theequation}{B.\arabic{equation}}
\begin{center} {\bf Appendix B: Proof of Theorem
\ref{thm:noisy_additive_LB}} \end{center}  \setcounter{equation}{0}

We derive a necessary condition on the number of tests $T$ for the additive noise model in (\ref{eq,additive_noise}) based on the Fano bound in Theorem \ref{thm,LB}. All we need is to upper bound the mutual information expression $I(X_{{\cal S}^1};X_{{\cal S}^2},Y)$ when the test outcome $Y$ is given by (\ref{eq,additive_noise}). From (\ref{eq:first_entropy}) and (\ref{eq:second_entropy}) we see that
\begin{align}
I(X_{{\cal S}^1};X_{{\cal S}^2},Y)&=\frac{i}{K\ln 2}\left(1-\frac{1}{K}\right)^K(1-q)+O\left(\frac{1}{K^2}\right)+\frac{1}{\ln 2}\left(1-\frac{1}{K}\right)^K (1-q)\nonumber\\
&-\frac{1}{\ln 2}\sum_{j=2}^{\infty}\frac{1}{j(j-1)}\left(1-\frac{1}{K}\right)^{K+i(j-1)}(1-q)^j-\left(1-\frac{1}{K}\right)^K q\log\frac{1}{q}
\end{align}
Combining terms using (\ref{eq:log_simplification}) and simplifying we get
\begin{align}
I(X_{{\cal S}^1};X_{{\cal S}^2},Y)&=\frac{i}{K\ln 2}\left(1-\frac{1}{K}\right)^K(1-q)+O\left(\frac{1}{K^2}\right)\nonumber\\
&+\frac{1}{\ln 2}\left(1-\frac{1}{K}\right)^K \sum_{j=2}^\infty \frac{1}{j(j-1)}(1-q)^j\left(1-\left(1-\frac{1}{K}\right)^{i(j-1)}\right)\nonumber\\
&\leq\frac{i}{K\ln 2}\left(1-\frac{1}{K}\right)^K(1-q)+O\left(\frac{1}{K^2}\right)\nonumber\\
&+\frac{1}{\ln 2}\left(1-\frac{1}{K}\right)^K\left(1-q+q\ln\frac{1}{q}\right)
\end{align}
where in the last inequality we made use of the fact that
\[
\sum_{n=1}^\infty \frac{x^n}{n(n+1)} = x + (x-1)\log\left(\frac{1}{1-x}\right)
\]
Since the lower bound in \ref{thm,LB} has to be satisfied for every $i$, then letting $i=K$ we have that
\begin{align}
I(X_{{\cal S}^1};X_{{\cal S}^2},Y)&=I(X_S,Y)\leq \frac{1}{\ln 2}\left(1-\frac{1}{K}\right)^K\left(2(1-q)+q\ln\frac{1}{q}\right)+O\left(\frac{1}{K^2}\right)
\end{align}
Replacing in (\ref{eq:lb_equation}) we get the result in Theorem \ref{thm:noisy_additive_LB}.

\renewcommand{\theequation}{C.\arabic{equation}}
\begin{center} {\bf Appendix C} \end{center}  \setcounter{equation}{0}
In this appendix we extend the result of Theorem \ref{thm:main_result} to the case where both $N$ and $K$ are allowed to scale simultaneously such that $K=o(N)$. We prove Theorems \ref{thm:noisless_appdx}, \ref{thm:additive_general} and \ref{thm:dilution_appdx}.

We will now establish a lower bound for the error exponent.
First, we begin with a simple lemma.
\begin{lemma} \label{thm:errexp}
$$
E_o(\rho) \geq \rho I(X_{\cs^1}; Y \mid X_{\cs^2}) - {\rho^2\over 2} \max_{0\leq \psi < 1}\left | (E_o(\psi))''\right |$$
\end{lemma}
\begin{proof}
We first note that from the Lagrange form of the Taylor Series expansion (essentially an application of the mean value theorem) we can write $E_o(\rho)$ in terms of its first derivative evaluated at zero and a remainder term, i.e.,
$$
E_o(\rho) = E_o(0) + \rho E_o'(0) + {\rho^2 \over 2} (E_o)''(\psi)
$$
for some $\psi \in [0,\rho]$.
We have already shown that $E_o(0)=0$ and $E_o'(0) = I(X_{\cs^1}; Y \mid X_{\cs^2})>0$. Consequently, we now lower bound $E_o(\rho)$ by taking the worst-case second derivative, i.e.,
$$
E_o(\rho) \geq \rho I(X_{\cs_1}; Y \mid X_{\cs_2}) -{\rho^2 \over 2} \max_{\psi \in [0,\rho]}|(E_o)''(\psi)|
$$
\end{proof}

We are left to bound the second derivative. Recall that, $E_o(\r)$ is defined as
\begin{align}
E_o(\r) = -\log\sum_{Y\in\{0,1\}}\sum_{X_{{\cal S}^2}}\left[\sum_{X_{{\cal S}^1}}Q(X_{{\cal S}^1}) p(Y,X_{{\cal S}^2}|X_{{\cal S}^1})^{\frac{1}{1+\rho}}\right]^{1+\rho}
\end{align}
For notational convenience we define:
$$
\br \models \br(Y,X_{{\cal S}^2},X_{{\cal S}^1} ) = p(Y,X_{{\cal S}^2}|X_{{\cal S}^1})^{\frac{1}{1+\rho}}
$$
Note, that $\br$ is a function of $Y,\,X_{{\cal S}^2},\,X_{{\cal S}^1}$ but we suppress these dependencies since we do not make use of this in our calculations other than to note that $0 \leq \br \leq 1$. Let,
$$
\fr \models E(\br) = \sum_{X_{{\cal S}^1}}Q(X_{{\cal S}^1}) p(Y,X_{{\cal S}^2}|X_{{\cal S}^1})^{\frac{1}{1+\rho}}
$$
Similarly, $\fr$ is still a random variable since it is dependent on $Y,X_{{\cal S}^2}$ but again this dependency is not used in much of our computation other than the fact that $\fr \in [0,\,1]$. Let,
$$
\gr \models \gr(Y,X_{{\cal S}^2}) = \fr^{1+\r}
$$
Again $\gr \in [0,\,1]$. Finally note that with these substitutions, we can rewrite, $E_o(\r)$ as:
$$
E_o(\r) = -\log\sum_{Y\in\{0,1\}}\sum_{X_{{\cal S}^2}}\gr(Y,X_{{\cal S}^2})
$$
Finally, let
$$
\ur = {\br \over E(\br)} ={p(Y \mid X_{\cs^1} X_{\cs^2})^{{1 \over 1+\rho}} \over \sum_{X_{\cs^1}} Q(X_{\cs^1}) p(Y \mid X_{\cs^1} X_{\cs^2})^{{1 \over 1+\rho}}}
$$
where the expectation is taken with respect to $X_{\cs^1}$. Note that the second equality follows by canceling out the common term $Q(X_{\cs^2})$ from both the numerator and denominator. Next, we have the following bound for the second derivative:
\begin{lemma}
If $E[u_\rho \log u_\rho]$ is a non-increasing function of $\rho$, then
$$
|(E_o(\rho))''| \leq \left | \sum_{X_{\cs^2}}\sum_{Y} g_{\rho}E(u_{\rho} \log^2(u_{\rho})) \right |
$$
where the expectation is taken with respect to $X_{\cs^1}$.
\end{lemma}
To prove this lemma we first establish the following propositions.
\begin{proposition}
\begin{align} \label{eq:secderbnd}
|E_o''(\r)| \leq \left |\frac{\sum_{Y,X_{{\cal S}^2}} \gr''}{\sum_{Y,X_{{\cal S}^2}} \gr} \right |
\end{align}
\end{proposition}
\begin{proof}
We rewrite the first and second derivatives of $E_0(\r)$ in terms of $\br,\fr,\gr$ and their first and second derivatives. Specifically, note that
$$
(E_o(\r))'' = -{\sum_{Y,X_{{\cal S}^2}} (\gr(Y,X_{{\cal S}^2}))'' \over \sum_{Y,X_{{\cal S}^2}} (\gr(Y,X_{{\cal S}^2}))} + \left({\sum_{Y,X_{{\cal S}^2}} (\gr(Y,X_{{\cal S}}))' \over \sum_{Y,X_{{\cal S}^2}} (\gr(Y,X_{{\cal S}^2}))}\right)^2
$$
Note that the second derivative of $E(\r)$ is negative for $\r > 0$. This follows from the fact that $E_0(\r)$ is a concave function (see \cite{gallager}). Furthermore, $\gr(\cdot)$ is a positive non-increasing function with $\sum_{Y,X_{\cs^2}}\gr=1$ at $\rho=0$. The proof now follows.
\end{proof}
Also note that the denominator in proposition \ref{eq:secderbnd} is bounded from below by $1/2$ since the mutual information is less than $1$:
\begin{align}
\sum_{Y,X_{{\cal S}^2}} \gr = 2^{-E_0(\rho)}\geq 2^{-I(X_{\cs^1};X_{\cs^2},Y)}\geq 1
\end{align}
We are now left to compute the second derivative for $\gr$. We have the following proposition:
\begin{proposition} \label{prp:props}
The following property holds:
$\gr'=-\gr E(u_{\rho} \log(u_{\rho}))$
\end{proposition}
\begin{proof}
\noindent
Note that,
$$
\fr' = - {1\over 1+\rho} E(\br \log \br),\,\,\gr' = \gr \log(\fr) + (1+\r) \fr^{\r} \fr'
$$
Note that $0 \leq \br \leq 1$. It follows that $\fr' > 0$ for positive $\r$. Simplifying the expression for $\gr'$ through substitution we obtain
$$
\gr'= \fr^{\r} (\fr \log(\fr) - E(\br \log(\br))= -\fr^{\r}E \left (\br \log {\br \over E(\br)}\right ) = - \gr E \left ({\br \over E(\br)} \log {\br \over E(\br)}\right )
$$
The proof now follows.

%
\end{proof}
We have the following result for the second derivative:
\begin{proposition} \label{prp:scnd}
$$
\gr'' = {\gr \over 1+\rho} E(u_{\rho} \log^2(u_{\rho})) + {\rho\gr \over 1+\rho} (E(u_{\rho} \log(u_{\rho})))^2
$$
It follows that if $E(\ur \log(\ur)$ is a decreasing function of $\rho$ then
$$
0 \leq \gr'' \leq \gr E(u_{\rho} \log^2(u_{\rho}))
$$
\end{proposition}
\begin{proof}
Note that,
\begin{align*}
\gr'' &= -\gr' E(u_{\rho} \log(u_{\rho})) -\gr {d \over d\rho} E(u_{\rho} \log(u_{\rho})) \\
=&\gr (E(u_{\rho} \log(u_{\rho})))^2-\gr {d \over d\rho} E(u_{\rho} \log(u_{\rho}))\\
=&\gr (E(u_{\rho} \log(u_{\rho})))^2-\gr E(u_{\rho}' \log(u_{\rho})+u_{\rho}')\\
=&\gr (E(u_{\rho} \log(u_{\rho})))^2-\gr E(u_{\rho}' \log(u_{\rho}))
\end{align*}
where the last equality follows from the fact that $E(\ur)=1$ and so $E(\ur')=0$. We now compute $\ur'$:
\begin{equation} \label{eq:dur}
\ur' = -{1 \over 1+ \rho} (\ur \log\br -  \ur E(\ur \log(\br))) = -{1 \over 1+ \rho} (\ur \log\ur - \ur E(\ur \log(\ur)))
\end{equation}
The first result now follows by direct substitution. To establish the second result we first note that the positivity of the expression is obvious since $\ur \geq 0$. For the upper bound we note from Equation~\ref{eq:dur}  that,
\begin{align*}
\phi_\rho={d \over d\rho} E(\ur \log(\ur)) = E(\ur' \log(\ur)+\ur') = E(\ur' \log(\ur))\\
=-{1 \over 1+ \rho} \left ( E(\ur \log^2\ur) - (E(\ur \log(\ur)))^2 \right )
\end{align*}
Now if $\phi_\rho \leq 0$  then it follows that $E(\ur \log^2(\ur)) \geq (E(\ur \log(\ur)))^2$ and the upper bound in the second part of the proposition would  follow by direct substitution.
\end{proof}

We have the following Lemmas for each of the different channels considered in the paper.
\begin{lemma}
Consider the noiseless channel. Then the worst-case second derivative $(E_o(\psi))''$ satisfies:
$$
|(E_o(\psi))''| = O\left({i \over K} \log^2 K\right).
$$
\end{lemma}
\begin{proof}
We first compute $\ur$ for the different cases. Note that for $X_{\cs^2}=0$ (i.e. all the components are zero) and $Y=0$ we have,
$$
P(Y=0 \mid X_{\cs^2}=0,X_{\cs^1}) = \left \{ \begin{array}{cc} 0 & \mbox{if}\,\,X_{\cs^1} \not = 0 \\1 & \,\,X_{\cs^1}= 0 \end{array} \right.
$$

$$
P(Y=0 \mid X_{\cs^2}=0) = (1-1/K)^i
$$
Therefore, $\ur = {1 \over (1-1/K)^i}$ or $\ur = 0$ for this case. Consequently, we immediately note that $E(\ur \log\ur)$ is independent of $\rho$. This implies that from Proposition~\ref{prp:props}
\[
\gr'' = \gr \log^2\frac{1}{(1-1/K)^i}
\]
which is a decreasing function of $\rho$. Hence the worst-case second derivative is realized at $\rho=0$.
Furthermore, note that in this case,
\[
(\gr)_{\rho=0} = P(Y=0, X_{\cs^2}=0)= (1-1/K)^K
\]

Suppose on the other hand $X_{\cs^2}=0$ but $Y=1$, then we get
\[
P(Y=1 \mid X_{\cs^1},X_{\cs^2}=0) = \left \{ \begin{array}{cc} 1 & \mbox{if}\,\,X_{\cs^1} \not = 0 \\0 & \,\,X_{\cs^1} = 0 \end{array} \right.
\]

\[
P(Y=0 \mid X_{\cs_1}=0) = 1-(1-1/K)^i
\]

Therefore, $\ur = \frac{1}{1-(1-1/K)^i}$ or $\ur = 0$ for this case. Consequently, we immediately note that $E(\ur \log(\ur))$ is independent of $\rho$. This implies that from Proposition~\ref{prp:props}
$$
\gr'' = \gr \log^2\frac{1}{(1-(1-1/K)^i)}
$$
which is a decreasing function of $\rho$. Hence the worst-case second derivative is realized at $\rho=0$. Furthermore, note that in this case,
$$
(\gr)_{\rho=0} = P(Y=1, X_{\cs^2}=0)= (1-1/K)^{K-i} (1-(1-1/K)^i)
$$

Note that $X_{\cs^2} \not = 0$ always results in $Y=1$ regardless of $X_{\cs^1}$ and hence $\ur$ for this case is always zero.

We are now ready to compute the second derivative based on the expression in Proposition~\ref{prp:scnd}. It follows by direct computation that,
$$
|(E_o(\psi))''| \leq |\sum_{X_{\cs^2},Y} (\gr)_{\rho=0} E(\ur \log^2\ur)| = O\left({i \over K} \log^2{K \over i}\right).
$$
\end{proof}

\subsection*{Proof of Theorem \ref{thm:noisless_appdx}}
We can readily prove the result in Theorem \ref{thm:noisless_appdx}.
Note that from the expression of $P(E_i)$ we have that,
$$
\sum_i P(E_i) \leq K \max_i P(E_i) \leq \max_i K \exp\left(-T\left(E_o(\rho) - {\rho \over T} \log\binom{N-K}{i} \binom{K}{i}\right)\right)
$$
Consequently, we need to ensure that,
$$
T E_o(\rho) \geq \rho \log\binom{N-K}{i} \binom{K}{i} + \log K
$$
Now using the lower bound for $E_o(\rho)$ we have
\begin{eqnarray*}
E_o(\rho) &\geq& \rho I(X_{\cs_2};Y \mid X_{\cs_1}) - {\rho^2 \over 2} |(E_o(0))''| \geq \rho I(X_{\cs_2};Y \mid X_{\cs_1}) - {\rho^2 \over 2} {i \over K} \log^2{K\over i} \\ & \geq & \rho {i\over K} - {\rho^2 \over 2} {i \over K} \log^2{K\over i}
\end{eqnarray*}
where the last inequality follows from the lower bound for the mutual information in the noiseless case (see Theorem~\ref{thm:avgPe_scaling}). Substituting this lower bound we obtain
$$
T\rho {i\over K} \left(1- {\rho\over 2} \log^2{K\over i}\right) \geq \rho \log\binom{N-K}{i} \binom{K}{i} + \log K
$$
By choosing $\rho= {1 \over \log^2(K/i)}$ and $T = 2K \log N \log^2(K)$ the inequality is satisfied.

\begin{lemma}
\label{lem:additive_Eo2}
Consider the additive noise channel. Then the worst-case second derivative $(E_o(\psi))''$ satisfies
\[
|(E_o(\psi))''| = O\left( {i \over K} \log^2\left(\frac{2}{q}\right)\right).
\]
\end{lemma}

\begin{proof}
We consider all the possible combinations for $Y$ and $X_{\cs^2}$.
\begin{itemize}
\item Case $Y=1$, $X_{\cs^2}=0$\\
\noindent 1) $E[\ur\log\ur]$ is monotone decreasing
\begin{proof}
$$
\ur = \left \{ \begin{array}{cc} \frac{q^{\frac{1}{1+\rho}}}{(1-\frac{1}{K})^iq^{\frac{1}{1+\rho}}+1-(1-\frac{1}{K})^i} & \mbox{if}\,\,X_{\cs^1} = 0 \\\frac{1}{(1-\frac{1}{K})^iq^{\frac{1}{1+\rho}}+1-(1-\frac{1}{K})^i}  & \,\,X_{\cs^1} \not = 0 \end{array} \right.
$$
To simplify notation define:
\begin{align}
a_1&\triangleq\left(1-\frac{1}{K}\right)^i\nonumber\\
q_{\rho}&\triangleq q^{\frac{1}{1+\r}}\nonumber\\
p&\triangleq \frac{a_1 q_\r}{a_1 q_\r+1-a_1}\nonumber
\end{align}
Hence,
\begin{align}
E[\ur\log\ur]&=p\log\frac{q_\r}{a_1 q_\r+1-a_1}+(1-p)\log\frac{1}{a_1 q_\r+1-a_1}\nonumber\\
&=-p\log\frac{1}{p}+p\log\frac{1}{a_1}+(1-p)\log(1-p)+(1-p)\log\frac{1}{1-a_1}\nonumber\\
&=-H(p)-p\log\left(\frac{a_1}{1-a_1}\right)+\log\frac{1}{1-a_1}
\end{align}
$p$ is a monotonically increasing function of $\r$. Hence, taking the derivative w.r.t. $p$
\begin{align}
\frac{d}{dp}E[\ur\log\ur]=\log\frac{p(1-a_1)}{a_1(1-p)}=\log q_\r < 0
\end{align}
\end{proof}

\noindent 2) $E[\ur\log\ur]$ is positive
\begin{proof}
$E[\ur\log\ur]$ is monotone decreasing. At $p=1$, $H(p)=0$, hence
\begin{align}
E[\ur\log\ur]=p\log\frac{a_1}{1-a_1}+\log\frac{1}{1-a_1}>0
\end{align}
establishing its positivity for all $p$.
\end{proof}
Hence, $\gr'' \leq \gr E(u_{\rho} \log^2(u_{\rho}))$. Now we are left to compute $E(u_{\rho} \log^2(u_{\rho}))$. Note that the maximum is further achieved at $\rho=0$ since $E(u_{\rho} \log^2(u_{\rho}))$ is monotone decreasing. This is not hard to see since $\ur$ takes $2$ values which approach $1$ as $\rho$ increases. Hence, $E(u_{\rho} \log^2(u_{\rho}))$ can be expanded as the sum of two monotone decreasing terms. Replacing with the values of $\ur$ at $Y=1$, $X_{\cs^2}=0$ we see that
\begin{align}
E[\ur\log^2\ur]\leq 2 {i \over K} \log^2\left(\frac{2}{q}\right)
\label{eq:urlog2ur_additive}
\end{align}
and
\begin{align}
(\gr)_{\rho=0} = P(Y=1, X_{\cs^2}=0)=\left(1-\frac{1}{K}\right)^{K-i}\left(q+1-\left(1-\frac{1}{K}\right)^i\right)
\label{eq:gr_additive_1}
\end{align}

\item Case $Y=0$, $X_{\cs^2}=0$\\
In this case,
$$
\ur = \left \{ \begin{array}{cc} 0 & \mbox{if}\,\,X_{\cs^1} \not = 0 \\
\frac{1}{(1-\frac{1}{K})^i}  & \,\,X_{\cs^1} = 0 \end{array} \right .
$$
so $\ur$ is independent of $\r$. Hence,
\begin{align}
\gr''=-\gr' E[\ur\log\ur]=\gr E^2[\ur\log\ur]
\end{align}
Noting that the first derivative $\gr'=-\gr E[\ur\log\ur]=-\gr\log\frac{1}{(1-\frac{1}{K})^i}\leq 0$, it follows that the third derivative $\gr'''=\gr'\log^2\frac{1}{(1-\frac{1}{K})^i}\leq 0$. Hence, the maximum of $\gr''$ is achieved at $\r=0$ and
\begin{align}
\gr''=(1-q)(1-\frac{1}{K})^K\log^2 \frac{1}{(1-\frac{1}{K})^i}
\label{eq:gr_additive_2}
\end{align}
\end{itemize}
Combining \ref{eq:urlog2ur_additive}, \ref{eq:gr_additive_1}, and \ref{eq:gr_additive_2} Lemma \ref{lem:additive_Eo2} follows. This result implies the same sufficient condition obtained in Theorem \ref{thm:noisy_additive} for the additive noise channel and is stated in Theorem \ref{thm:additive_general}.
\end{proof}

\begin{lemma}
\label{lem:dilution_Eo2}
Consider a dilution channel with dilution probability $p$. Then the worst-case second derivative $(E_o(\psi))''$ satisfies:
$$
|(E_o(\psi))''| = O\left( \frac{i(1-p)}{K}\log^2 K\right).
$$
\end{lemma}
\begin{proof}
The proof consists of considering different cases and evaluating the expression in Proposition VII.3.  The proof requires elementary algebraic operations and we omit them for brevity. In all of the cases it turns out that the terms $E(u_\rho \log u_\rho)$ and $E(u_\rho \log^2 u_\rho)$ are bounded and so the problem boils down to computing the maximum value of $g_\rho$. Since $g_\rho$ is monotonically decreasing, its maximum value occurs at $\rho=0$. This is what we show here for the different cases.
\begin{itemize}
\item Case $Y=0, X_{\cs^2}=0, ||X_{\cs^1}||_1=\ell$
i.e., $X_{\cs^1}$ has $\ell$ ones.
\begin{align}
\ur(\ell)=\frac{p_{\r}^\ell}{\sum_{\ell=0}^i \binom{i}{\ell}(\frac{1}{K})^\ell(1-\frac{1}{K})^{i-\ell}p_{\r}^\ell}
\end{align}
where $p_{\r}=p^{\frac{1}{1+\r}}$. 
In this case
\[
(\gr)_{\rho=0} = P(Y=0, X_{\cs^2}=0) = \left(1-\frac{1}{K}\right)^{K-i}\left(1-\frac{1-p}{K}\right)^i
\]

\item Case $Y=0, X_{\cs^2}\not=0, ||X_{\cs^1}||_1=j$\\
This case parallels the previous case but replacing $p_{\r}^{\ell}$ with $p_{\r}^{\ell+j}$

\item Case $Y=1, X_{\cs^2}=0, ||X_{\cs^1}||_1=\ell$
\begin{align}
\ur=\frac{(1-p^\ell)^{\frac{1}{1+\r}}}{\sum_{\ell=0}^i\binom{i}{\ell}(\frac{1}{K})^\ell(1-\frac{1}{K})^{i-\ell}(1-p^\ell)^{\frac{1}{1+\r}}}
\end{align}
In this case, 
\[
(\gr)_{\rho=0} = P(Y=1, X_{\cs^2}=0) = \left(1-\frac{1}{K}\right)^{K-i}\left[1-\left(1-\frac{1-p}{K}\right)^i\right] =\frac{i(1-p)}{K} + O\left(\left(\frac{i}{K}\right)^2\right)
\]

\item Case $Y=1, X_{\cs^2}=j\ne 0, ||X_{\cs^1}||_1=\ell$\\
This case parallels the previous case replacing $1-p^\ell$ with $1-p^{\ell+j}$. Also
\begin{align}
(\gr)_{\rho=0} &= P(Y=1, ||X_{\cs^2}||_1=j) \nonumber\\ &=\left(\frac{1}{K}\right)^j\left(1-\frac{1}{K}\right)^{K-i-j}\left[1-p^j\left(1-\frac{1-p}{K}\right)^i\right]\nonumber\\
&\leq \left(\frac{1}{K}\right)^j(1-p^{j+1})
\end{align}
\end{itemize}

Now following similar arguments to the previous additive and noiseless cases,
\begin{enumerate}
\item Consider the first case and other cases could be handled similarly. We note that
\begin{align}
E[\ur\log\ur]&=\sum_{\ell=0}^{i}\binom{i}{\ell}\left(\frac{1}{K}\right)^\ell\left(1-\frac{1}{K}\right)^{i-\ell} \frac{p_{\r}^\ell}{(1-\frac{1-p_{\r}}{K})^i}\log\frac{p_{\r}^\ell}{(1-\frac{1-p}{K})^i}\nonumber\\
&=\sum_{\ell=0}^{i}\ell\binom{i}{\ell}(\frac{p_{\r}}{K})^\ell(1-\frac{1}{K})^{i-\ell} \frac{1}{(1-\frac{1-p_{\r}}{K})^i}\log p_{\r}+\log\frac{1}{(1-\frac{1-p_{\r}}{K})^i}\nonumber\\
&=\frac{ip_{\r}\log p_{\r}}{K(1-\frac{1-p_{\r}}{K})}+\log\frac{1}{(1-\frac{1-p_{\r}}{K})^i}=O(i/K).
\end{align}

\item Following a similar argument we can also show that for the $Y=0$, $X_{\cs^2}=0$ case
\begin{align}
E[\ur\log^2\ur]&=\sum_{\ell}\binom{i}{\ell}\left(\frac{1}{K}\right)^\ell\left(1-\frac{1}{K}\right)^{i-\ell} \frac{p_{\r}^\ell}{(1-\frac{1-p_{\r}}{K})^i}\log^2\frac{p_{\r}^\ell}{(1-\frac{1-p_{\r}}{K})^i}\nonumber\\
&=\frac{(\frac{ip_{\r}}{K})^2\log^2 p_{\r}}{(1-\frac{1-p_{\r}}{K})^2}+i^2\log^2\frac{1}{(1-\frac{1-p_{\r}}{K})}
\end{align}
Similarly, for $Y=1$, $X_{\cs^2}=0$, at $\r=0$
\begin{align}
E[\ur\log^2\ur]&=\sum_{\ell=0}^{i}\binom{i}{\ell}\left(\frac{1}{K}\right)^\ell\left(1-\frac{1}{K}\right)^{i-\ell} \frac{1-p^\ell}{1-(1-\frac{1-p_{\r}}{K})^i}\log^2\frac{1-p^\ell}{1-(1-\frac{1-p}{K})^i}\nonumber\\
&=O\left(\log^2\frac{K(1-p)}{i}\right).
\end{align}
Combining the results for the various cases establishes the result in Lemma \ref{lem:dilution_Eo2} and Theorem \ref{thm:dilution_appdx}.
\end{enumerate}
\end{proof}

\section{Acknowledgments}
The authors would like to thank A. Sahai at UC Berkeley and N. Ma at Boston University for very useful discussions, and the anonymous referees for
comments that improved the presentation.
\begin{IEEEbiographynophoto}{George K. Atia}
(S'01--M'04) received the B.Sc. and M.Sc. degrees from Alexandria University, Egypt, in 2000 and 2003, respectively, and the Ph.D. degree from Boston University, MA, in 2009, all in electrical and computer engineering.

He joined the Department of Electrical and Computer Engineering at the University of Illinois at Urbana-Champaign in Fall 2009 where he is currently a postdoctoral research associate in the Coordinated Science Laboratory. His research interests include statistical signal processing, wireless communications, information and decision theory.

Dr. Atia is the recipient of many awards, including the Outstanding Graduate Teaching Fellow of the Year Award in 2003$-$2004 from the Electrical and Computer Engineering Department at Boston University, the 2006 College of Engineering Dean's Award at the BU Science and Engineering Research Symposium, and the best paper award at the International Conference on Distributed Computing in Sensor Systems (DCOSS) in 2008.
\end{IEEEbiographynophoto}

\begin{IEEEbiographynophoto}{Venkatesh Saligrama}
(SM'07) Venkatesh Saligrama is a faculty member in the Electrical and Computer Engineering Department at Boston University. He holds a PhD from MIT. His research interests are in Statistical Signal Processing, Statistical Learning, Video Analysis, Information and Decision theory. He has edited a book on Networked Sensing, Information and Control. He has served as an Associate Editor for IEEE Transactions on Signal Processing and Technical Program Committees of several IEEE conferences. He is the recipient of numerous awards including the Presidential Early Career Award(PECASE), ONR Young Investigator Award, and the NSF Career Award. More information about his work is available at http://blogs.bu.edu/srv
\end{IEEEbiographynophoto}
\end{document}